\documentclass{elsart}
\usepackage{graphics,epsf}
\usepackage{rotating}
\usepackage{bbm}
\usepackage{revsymb}

\newcommand{\G}{\gamma}

\newcommand{\EPEM}{e^+e^-}

\newcommand{\A}{\alpha}
\newcommand{\ZA}{Z\alpha}
\newcommand{\BE}{\begin{equation}}
\newcommand{\EE}{\end{equation}}

\makeatletter
\def\lesssim{\mathrel{\mathpalette\vereq<}}
\def\vereq#1#2{\lower3pt\vbox{\baselineskip1.5pt \lineskip1.5pt
\ialign{$\m@th#1\hfill##\hfil$\crcr#2\crcr\sim\crcr}}}
\def\gtrsim{\mathrel{\mathpalette\vereq>}}
\makeatother

\begin{document}
\runauthor{Baur, Hencken, Trautmann}
\begin{frontmatter}
\title{
Electron-Positron Pair Production in Relativistic 
Heavy Ion Collisions}
\author[FZJ]{Gerhard Baur, }
\author[UBA,ABB]{Kai Hencken,}
\author[UBA]{and Dirk Trautmann}
\address[FZJ]{Forschungszentrum J\"ulich, J\"ulich, Germany}
\address[UBA]{Universit\"at Basel, Basel, Switzerland}
\address[ABB]{ABB Corporate Research, Baden D\"attwil, Switzerland}
\begin{abstract}
\label{section:abstract}

In recent years, a large number of papers have appeared that dealt with 
$\EPEM$ pair production in heavy ion collisions at
high energies. To a large extent these studies were motivated 
by the existence of
relativistic heavy ion accelerators all over the world.
There pair production can be studied in so called 
``ultra-peripheral collisions'', where the ions do 
not come close enough to interact strongly with each other. 
Various different methods have been used and it is the purpose of this review
to present a unified picture of the present status of the field.
The lowest order Born result has been known for more than 
seven decades. The interest and focus
is now on higher order effects for values of $Z\alpha 
\lesssim 1$, where $Z$ is the charge 
number of the ion. 
A similar 
problem appears for the Bethe-Heitler process, the 
production of $\EPEM$ pairs in photon-nucleus collisions.
It was solved essentially some five decades ago by Bethe and Maximon.
The result of Bethe and Maximon can also be
recovered by summing over a class of Feynman diagrams to infinite
order.
These results can be used  for 
a study of Coulomb corrections in nucleus-nucleus collisions.
Indeed, the major part of these corrections have a structure
closely related to the Bethe-Maximon solution.
There are additional terms which give a small contribution to the 
total cross section
at high energies. Their importance can be enhanced by concentrating
on small impact parameters. 
An interesting exact solution of the one-particle Dirac
equation in the high-energy limit was found independently by several
authors. This led to some discussion about the interpretation of
these results within QED and the correct regularization necessary to
get the correct result. The dust of  previous debates
has settled and, indeed, a consistent picture has emerged.
Another interesting higher order effect is multiple pair
production, which we also discuss. We compare experimental results
obtained recently at RHIC for free and bound-free pair 
production with theoretical results. We also make some more remarks on the 
physics of strong electric fields of longer duration. A new field is opened up
by ultra-intense laser pulses. We argue that due to the short interaction time 
in ultraperipheral heavy ion pair production can
be well understood in the frame of QED perturbation theory. 
\end{abstract}

\end{frontmatter}

\tableofcontents

\section{Introduction, overview  and purpose}
\label{section:genintro}

Quantum electrodynamics (QED) is the best-tested theory we have
\cite{Phys.Rev.76.790,Phys.Rev.76.749,Phys.Rev.76.769,Schwinger:1958}.
It is a quantum field theory which describes the 
gauge invariant interaction of charged particles with photons.
With this theory one is able to describe high precision experiments 
like the Lamb shift and the magnetic moments of the
electron and muon with great accuracy.
In the present review we are dealing with the fundamental process of
particle production in high energy scattering processes. Experiments
up to now have been not very accurate in their results.
Also, the theoretical precision was not very high. This is not due to
principal problems of the underlying theory, the QED. It is more related to
technical difficulties, as we shall see in the following.
An especially interesting process is the production of
electron-positron pairs under various conditions.

Positrons were first observed in cosmic
ray interactions in 1932 \cite{Anderson:1933xx}.
The theoretical calculation of $e^+e^-$ pair production
in fast nucleus-nucleus collisions goes back to the early days
of QED \cite{Landau:1935aa}.
The full result in lowest order 
perturbation theory (Born approximation) is due to Racah
\cite{NUCIA.14.93}. These results are also discussed in 
\cite{PRPLC.15.181}.
In those calculations the pair production is due to the collision 
of a pair of (virtual) photons.
The theoretical, as well as, experimental interest in 
these processes was spurred in the last decades by the 
relativistic heavy ion facilities AGS at Brookhaven, SPS 
at CERN and more recently RHIC at Brookhaven  and LHC at CERN.
It was 
realized that the old lowest order answers are no longer sufficient
for the strong Coulomb field surrounding the heavy ions at these high
energies. There are (new) sizeable and interesting higher order effects. 

The subject of this review is  precisely the situation envisaged in
these early works of the thirties of the past century, 
i.e. pair production in high energy nucleus-nucleus collisions.
The theoretical treatment now concentrates on higher order
effects. One type of effects is the 
influence of higher orders on the one-pair production.
Very similar ``Coulomb corrections'' have been discussed
already before in pair production by a single (real) photon in the
Coulomb field of a nucleus. As we will see a lot of the experience
gained there can also be applied to the case of two heavy ions.
Another --- new and more spectacular --- effect is the production of 
many pairs in a single collisions, clearly a higher order effect.
Various groups have been working on these problems in the past two 
decades using various methods and  approximations. 
The main purpose of this review is to 
discuss these various methods and to establish the 
relationship of the different approaches.
Since QED is a well established theory,
such a connection is to be expected.
  
Before dealing specifically with our subject of relativistic heavy ion
collisions, it is appropriate to put the present subject of 
$\EPEM$ production in relativistic nucleus nucleus collisions
into a wider perspective:
slow collisions of heavy ions of charge numbers $Z_1$ and $Z_2$
are a way to create, at least for a short time,
atoms with a charge $Z_{\rm united \; \rm atom} =Z_1 + Z_2$.
It is of special interest to 
study collisions where  the 
sum of the charges   $Z_{\rm united \; \rm atom}>137$ (173).
In these collisions so called ``overcritical fields'' are reached for
some time where spontaneous $\EPEM$ production is possible,
see \cite{Rafelski:1978} or the detailed review of this 
field in \cite{GreinerStrongField}.
The supercritical field is reached for a charge $Z=137$ 
in the case of a point nucleus, where one may say that the  
$1s_{1/2}$-level dives into the negative energy continuum.
For extended nuclei with a finite radius, this limit 
is reached at about $Z=173$ \cite{Rafelski:1978}. 

\begin{figure}
\centering
     \includegraphics[width=6cm]{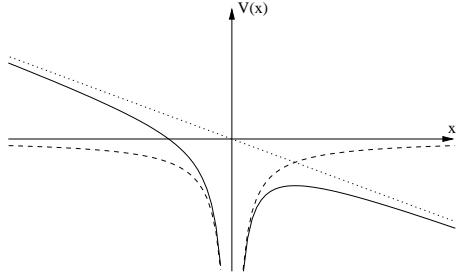}
\caption{Pair production in a strong static field $E$. The total
  potential is given by the potential energy (solid line) of the
  electron in the binding potential $V(x)$ (dashed line) and in the
  static electric potential $eE$. We may consider the electron bound
  by a potential of depth $\sim 2 m c^2$. In the Schwinger process
  the electron tunnels through the barrier and an $\EPEM$ pair is
  created, see \protect\cite{ItzyksonZ80}.
}
\label{fig:schwingermechanism}
\end{figure}

The paradigm of $\EPEM$ pair production is the Schwinger mechanism:
the production of pairs  in a constant
spatially uniform electric field \cite{Phys.Rev.82.664}, see
Fig.~\ref{fig:schwingermechanism}.
To reach a sizeable rate for spontaneous production of $\EPEM$ pairs
a field strength of the order of or above 
a certain critical field strength
$E_c$ is required. The rate of (one) pair production per unit volume
and time for a given constant field $E$ is given by \cite{ItzyksonZ80}
\BE
\frac{d^4n_{\EPEM}}{d^3xdt} \sim \frac{c}{4 \pi^3 \lambdabar^4}
\exp(-\pi\frac{E_c}{E}),
\EE
where the critical field strength is given by 
\begin{equation}
E_c= \frac{m^2c^3}{e \hbar} \simeq 1.3 \times 10^{16} V/cm.
\end{equation}
The electron mass is denoted by $m$ and its Compton wavelength
is given by $\lambdabar=\frac{\hbar}{m c}$.
This formula clearly shows the non-perturbative nature of the problem:
there is no power series expansion of the $\EPEM$ production rate in
terms of the electric field strength $E$. 
With new developments in laser physics 
it seems that this field strength can be reached, e.g., at the FEL at
DESY and the interest has 
grown enormously, both in theory and experiment \cite{Ringwald:2003er,Ringwald:2001ib}.
In contrast to relativistic heavy ion collisions, the created fields
in this case are 
of a comparatively long duration. This is made more quantitative 
in \cite{Phys.Rev.D2.1191} 
and will be dealt with in Sec. \ref{chap:fastslow}.

\begin{figure}
\centering
     \includegraphics[width=4cm]{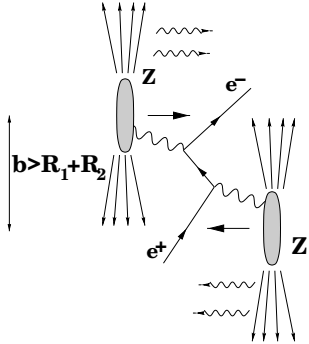}
\caption{The electromagnetic fields of two heavy ions in
an ultraperipheral collision. These fields can be decomposed into a
spectrum of quasireal photons (Weizs\"acker-Williams approximation). The
collision of two photons gives rise to a dilepton pair. The collision of 
photons radiated from each nucleus are a copious source 
of $\EPEM$ pairs.
}
\label{fig:hioncollision}
\end{figure}

The main purpose of relativistic heavy ion collisions 
is the production and observation of the quark gluon plasma
which yields information about QCD at high temperature
and densities. It has gradually been
realized over the last decades that the very strong electromagnetic fields
present in distant heavy ion collisions are also
useful: they provide a strong flux of equivalent photons for
photon-hadron and photon-photon physics with hitherto unknown energies.
The term ``ultraperipheral collisions'' (UPC) has been coined to 
describe collisions where the impact parameter is larger than the 
sum of the two nuclei: in this case only electromagnetic interactions
occur between the ions.
One special case of photon-photon physics in UPCs is $\EPEM$
production, see Fig.~\ref{fig:hioncollision}.
Due to the smallness of the electron mass $\EPEM$ pair production 
is very copious,
it is of the order of kilobarns under RHIC and LHC conditions.
The physics of ``ultraperiphal collisions'' is reviewed in
\cite{Krauss:1997vr,Baur:1998ay,PRPLC.364.359,Bertulani:2005ru} and 
\cite{PRPLC.163.299}.
A recent workshop at ECT*/Trento,Italy was devoted to the subject of
UPC \cite{Baltz:2007hw}.
The status of $\EPEM$ pair production was also
discussed there, see \cite{ect07website}.

Let us make some order of magnitude estimates of the 
fields acting in the case of relativistic heavy ion collisions:
the electromagnetic fields encountered 
in these collisions are extremely strong,
they are of the order of 
\BE
 E_{\rm max}\simeq \frac{Z e \gamma}{b^2},
\label{eq:maxfield}
\EE 
where we assume that the ions move on straight line 
trajectories with impact parameter
$b$ and $\gamma$ is the Lorentz factor in the collider system. 
With the values of $Z=79$, $b=b_{min}=15fm $ and $\gamma=100$, which are 
appropriate for RHIC, we find
a maximum field strength of  $E_{\rm max}= 4.9 \times 10^{16} V/cm$. 
For LHC conditions ($Z=82, b_{min}=15fm, \gamma=3000$)
we get $E_{\rm max}= 1.5 \times 10^{18} V/cm$.
These are the values viewed in the collider frame.
Viewed from the other nucleus the corresponding Lorentz factor is
$\gamma_{ion}=2 \gamma^2 -1$ and the electric fields are even 
larger by a factor of $\gamma$.

These field strengths are comparable to the critical field $E_c$
for the Schwinger mechanism.
However, these fields act only for a very short time 
\BE
\Delta t \simeq \frac{b}{\gamma v},
\label{eq:deltat}
\EE
where $v \sim c$ is the ion velocity. 
For $b=\lambdabar\simeq 386$fm one obtains 
the  values of $\Delta t \simeq 10^{-23}s$ for RHIC and 
$\Delta t \simeq 3\times 10^{-25}s$ for LHC conditions, respectively.
The momentum transfer $\Delta p$ given to a test particle with a
charge $e$ along a path with impact parameter $b$ is independent of 
$\gamma$. For $v \sim c$ and $b=\lambdabar$ it is only modest
\BE
\Delta p \sim e E_{max} \Delta t \sim \frac{Z e^2}{bv} \sim  Z\alpha
m c.
\EE
This small value gives a strong hint that the use of perturbation
theory is still appropriate
in contrast to the  situation of the Schwinger mechanism mentioned above.
Surely there are higher order effects, but they can 
be dealt within higher order perturbation theory, see, e.g.,
\cite{NUPHA.A729.787}.
One goal of this review is the explanation of the results for the effects 
beyond lowest order perturbation theory. An obvious question is about 
the transition from the slow to the fast collisions.

In section \ref{section:behei} we recall a simpler but intimately 
related problem: the Bethe-Heitler process, i.e. the photoproduction of $\EPEM$
pairs in the nuclear Coulomb field. 
It shows already many of the characteristic aspects of the
heavy ion case. Higher order effects were treated in 
\cite{Bethe:1954bmd,PhysRev.93.788}
by using the Sommerfeld-Maue wave functions: these
wave functions are appropriate high energy solutions of the Dirac 
equation and thus take higher order effects
into account to all orders. This is a textbook example (see e.g. 
\cite{Landau:1986aa}). It was shown in
\cite{Phys.Rev.D57.4025} how this approach is related to
the usual Feynman graph technique.

In section \ref{section:fdia} we come back to the case of 
ultraperipheral heavy ion collisions.
We first review the Feynman graph approach to pair production 
in relativistic heavy ion collisions.
The lowest order (Born approximation) cross section was given by 
Landau and Lifshitz \cite{Landau:1935aa}  and Racah 
\cite{NUCIA.14.93} in the thirties of the past
century. In contrast to the photoproduction on a nucleus
one has a well-defined impact parameter in 
heavy ion collisions and impact-parameter
dependent probabilities have been studied as well.
Higher order effects were studied in the last decades, 
mainly by the russian schools.
We note here that due to the high mass of the projectile
bremsstrahlung pair production is negligible at the high energies, 
as explained, e.g., in \cite{Landau:1986aa}, see also \cite{Meier:1997mp}.
Bremsstrahlung pair production is relevant for lower beam energies, where
the (timelike) bremsstrahlungs-photon is converted to an $\EPEM$ pair. 

\begin{figure}
\centering
     \includegraphics[width=4cm]{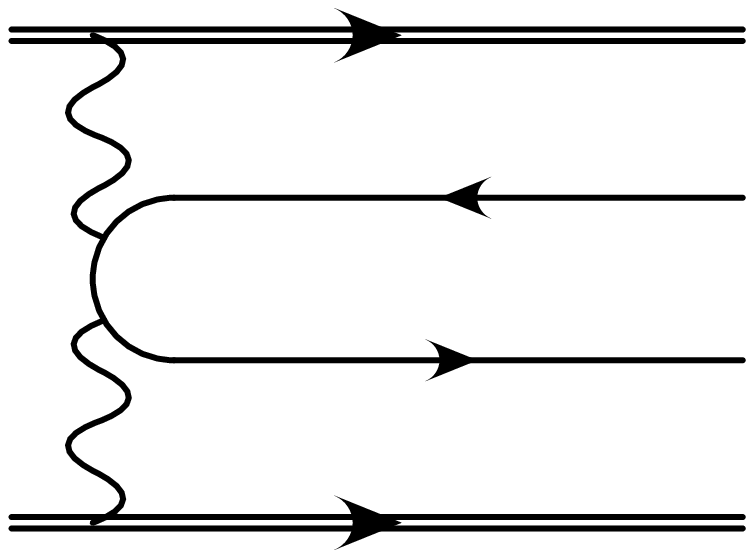}(i)
     \includegraphics[width=4cm]{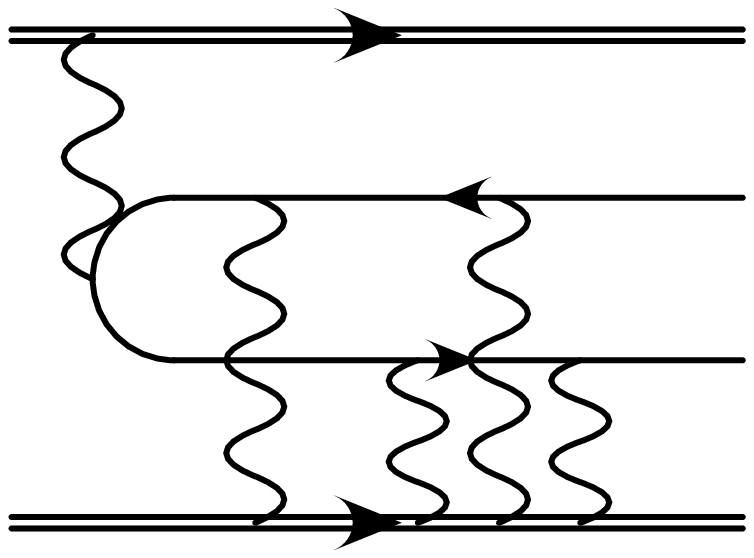}(ii)\\
     \includegraphics[width=4cm]{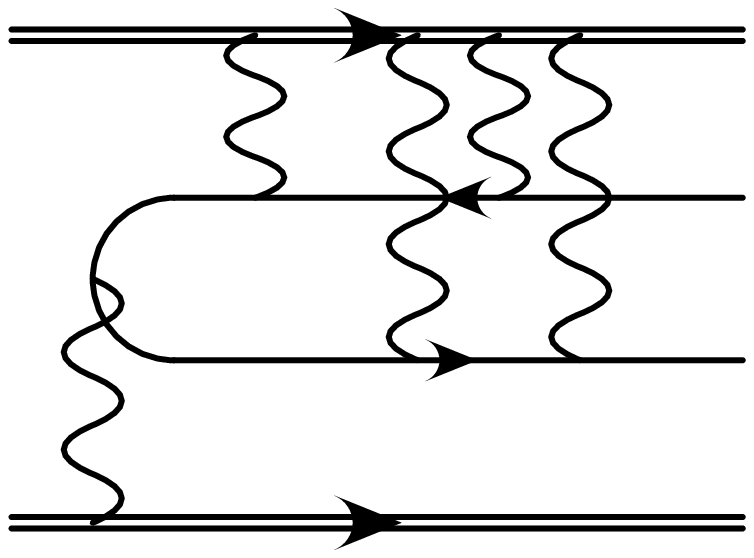}(iii)
     \includegraphics[width=4cm]{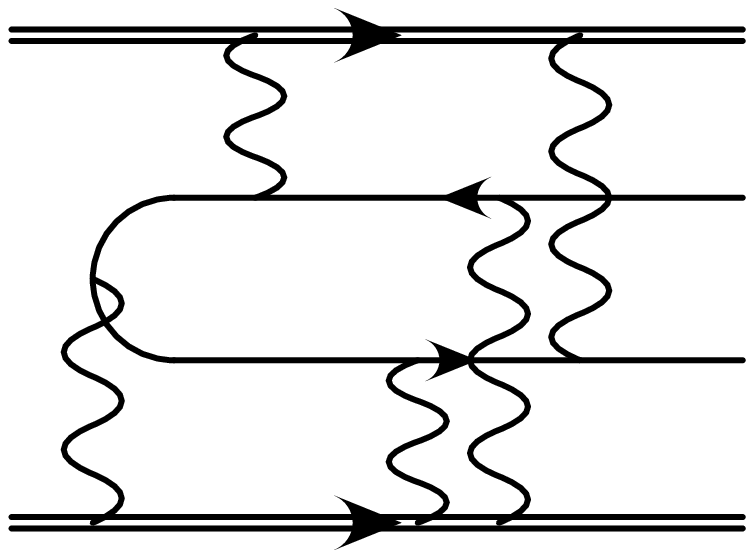}(iv)
\caption{$\EPEM$ pair production is classified by the number of
photons attached to each ion. We distinguish for classes
(i) $n=n^\prime=1$ (ii) $n=1$, $n^\prime>1$ (iii) $n>1$, $n^\prime=1$
and (iv) $n>1$, $n^\prime >1$.
}
\label{fig:classesitoiv}
\end{figure}

As we will see in the following we have an important classification 
of the terms:
We denote the number of photons attached to nucleus 1 by $n$,
to nucleus 2 by $n^{'}$ and distinguish four classes, see
Fig.~\ref{fig:classesitoiv}:\\
(i) $n=1$ and $n^\prime=1$\\
This is the lowest order case as treated by Landau and Lifshitz
\cite{Landau:1935aa} and Racah \cite{NUCIA.14.93}. The cross section 
increases with the Lorentz factor $\gamma$ like
$(\log \gamma)^3$ and is the dominant term at high energies.\\
(ii)  $n=1$ and $n^\prime>1$\\
(iii) $n>1$ and $n^\prime=1$\\
These terms are the main correction and 
well understood as we will show. The problem is essentially the same
as the one for the Bethe-Heitler process, and it is solved in an
analogous way. 

The case, which is specific to heavy ion collisions and which contains
all the problems is the last one:\\
(iv) $n>1$ and $n^\prime>1$\\
This case cannot be considered to be solved, but at least 
it is reassuring that one can show that
it is only a small contribution to the cross section.
This classification is directly evident in the Feynman diagram
approach of Sec. \ref{section:fdia},
but it also plays the major role in the approaches of Sec.
\ref{section:serboetal}. When the ion velocity decreases or for small
impact parameter the Feynman diagrams of class (iv) will become more important.
So collisions at intermediate and small energies become
increasingly difficult to treat theoretically. Molecular orbital approaches 
will then become relevant. This is quite analogous to the situation in
atomic physics for the excitation of atoms by ions in ion-atom collisions. This is an
interesting topic but it is outside the scope of this review.

In Sec. \ref{section:mclerran} an analytical solution to the 
(one-particle) time dependent two-center Dirac equation 
in the high energy limit
is discussed. The authors of \cite{Phys.Rev.C58.1679,nucl-th/0101024,Phys.Rev.A57.1849,Phys.Rev.C59.2753}
were led to the (erroneous) conclusion that the 
cross section corresponding to the exact solution is the same as the 
cross section in the Born approximation. But this is in 
obvious contradiction to the special case of the Bethe-Maximon corrections
to the Bethe-Heitler formula.
There are two aspects in this work:
one is related to the interpretation of the one-particle Dirac equation
to the corresponding (many-body) Feynman diagrams describing pair production.
We will show how the two can be related, see,
e.g. \cite{Aste:2001te,Phys.Rev.C71.024901}. The other aspect has to do with the proper regularization
of the expression, which was used by
\cite{Phys.Rev.C58.1679,Phys.Rev.A57.1849,Phys.Rev.C59.2753,Phys.Rev.A59.1223}
in the high energy (sudden) limit. 

In Sec. \ref{section:serboetal}  
we discuss the work of Ivanov, Schiller and Serbo \cite{Ivanov:1998ru}
and Lee and Milstein \cite{Phys.Rev.A65.022101}.
The latter work is based on the analytical solution of the Dirac
equation in Sec. \ref{section:mclerran}.
In Sec. \ref{section:multred} 
we discuss some aspects of class (iv): two- or more photons 
emerge from either nucleus: in addition to the 'Coulomb corrections',
which are small, there is a new genuine 
higher order effect: the multiple pair production in a single 
collision.
Another higher order effect (belonging essentially to 
classes (ii) and (iii)) is bound-free pair production:
the electromagnetic interaction of the ion
(to all orders) with the 
electron leads to a bound state ($K$-,$L$-shell,\ldots capture).
This is a small but interesting fraction 
of the total pair production, the cross section scales
only with $\log(\gamma)$, instead of $(\log(\gamma))^3$ for  free pairs.
In Sec. \ref{section:fastslow}  
we make some remarks about the transition from the 
non-perturbative slow to the perturbative fast collisions.
This is a difficult question. We discuss especially the results 
of \cite{Phys.Rev.D2.1191}. An interpolation between the two
limits is achieved analytically for the case of spatially constant
electric fields. 
In Sec. \ref{section:compa} we give a comparison of theory to experiment:
in the last five years since our previous review \cite{PRPLC.364.359}
RHIC has come into operation. We compare the experimental results 
of STAR on $\EPEM$ pair production in UPC \cite{Adams:2004rz} to
lowest order QED calculations and discuss the results of 
bound-free pair production at RHIC. Accelerator physicists 
are anxious to test the theory of bound-free pair production
because this process is crucial for the operation of
the forthcoming LHC(Pb-Pb). The bound-free
capture process will be the ultimate limit for the Pb-Pb luminosity.
Conclusions and an outlook are given in Sec. \ref{section:conclusion}.

Throughout this review we use $\hbar=c=1$ and $\alpha=e^2$, unless it
seems of some pedagogical value to include these constants explicitly.
We also assume relativistic high energy collisions, which are
characterized by $\gamma\gg 1$. We assume that even $\log(\gamma)\gg
1$. This is well fulfilled for RHIC and LHC conditions, less so for
AGS and SPS energies.

\section{Lepton pair production by a high energy
photon in the strong Coulomb field of a heavy nucleus}
\label{section:behei}

In this chapter we recall the well known theoretical treatment of the Bethe-
Heitler process, the $\EPEM$pair production by a photon 
in the Coulomb field 
of a nucleus. This is simpler than the problem of 
pair production in the field of two nuclei, but shows already
many of the important aspects of the nucleus-nucleus case. 
It corresponds to the cases (i), (ii) and (iii) (see Sec. \ref{section:genintro}), 
where only one 
photon is attached to one of the nuclei. Case (iv) is the one, which
is specific to the nucleus-nucleus problem.

The Born approximation result (case (i)) is derived in textbooks
using the well established tools of QED, see e.g. \cite{Landau:1986aa}. 
The celebrated Bethe-Heitler result for the total cross section of the
process   
\BE
\gamma + Z \rightarrow e^+ + e^- + Z
\EE
in the case of no screening is given in the high energy limit by
\BE
\sigma_1=\frac{28}{9} \frac{\alpha ^3 Z^2}{m^2}(\log{\frac{2 \omega}{m}}-
\frac{109}{42}),
\label{eq:behei}
\EE 
where the photon energy is denoted by
$\omega$. The Coulomb field of the nucleus is treated as an external field.
It is a characteristic feature that the cross section rises
logarithmically with the photon energy. 
Screening due to the atomic electrons
leads to a constant cross section for high energies. This 
screening effect
is of little interest for us here,
as we are mainly interested in the case of bare ions at the colliders.

The higher order effects in pair production at high photon 
energies and also  for large energies of  electron and positron 
compared to their rest mass,
were studied by Bethe, Maximon and Davies \cite{Bethe:1954bmd,PhysRev.93.788}.

Their results are also described in detail in many textbooks 
(see especially \cite{Landau:1986aa}). For the derivation of the 
cross section these authors used Sommmerfeld-Maue wave functions 
for the electron and the positron, respectively. 
These wave functions are approximate solutions  of the Dirac equation 
in the nuclear Coulomb field valid for high energies and small
scattering angles. 
Thus they include, in a certain way, higher order effects up to
all orders. They are given for 
the electron with energy  $\varepsilon_-$
and momentum $\overrightarrow{p}_-$ by:

\BE
\psi^{(-)}_{\varepsilon_-p_-} = \frac{N_-}{\sqrt{2\varepsilon_-}} 
e^{i \overrightarrow{p}_- \cdot \overrightarrow{r}}
(1-i\frac{\overrightarrow{\alpha} \cdot \nabla}{2\varepsilon_-})
F(-i Z \alpha,1,-i (p_-r + \overrightarrow{p}_- \cdot \overrightarrow{r})) u(p_-)
\EE
with the normalization
\BE
N_- = \exp({\frac{\pi}{2} {Z\alpha }})  \Gamma(1+iZ\alpha)
\EE
depending on the $\Gamma$-function and 
the confluent hypergeometric function $F$. The Dirac spinor for the
electron in the free case is given by $u$ and $\overrightarrow \alpha$ are the Dirac matrices.

Similarly the wave function for the positron is given by:
\BE
\psi^{(+)}_{-\varepsilon_+p_+} = \frac{N_+}{\sqrt{2\varepsilon_+}} 
e^{-i \overrightarrow{p}_+ \cdot \overrightarrow{r}}
(1+i\frac{\overrightarrow{\alpha}\cdot \nabla}{2\varepsilon_+})
F(-i Z\alpha,1,i (p_+r + \overrightarrow{p}_+ \cdot \overrightarrow{r})) u(p_+)
\EE
with
\BE
N_+ = \exp({-\frac{\pi}{2} {Z\alpha}})  \Gamma(1-iZ\alpha).
\EE
Inserting these wave functions into the standard matrix element, 
differential and total cross section can be calculated analytically,
leading (after neglecting some small terms in the high energy limit) 
to the expression for the total cross section:
\BE
\sigma_1=\frac{28}{9} \frac{\alpha ^3 Z^2}{m^2}(\log{\frac{2 \omega}{m}}-
\frac{109}{42}-f(Z\alpha)),
\EE
where the Bethe-Heitler cross section, 
see Eq.~(\ref{eq:behei}) above is now modified by 
a term containing the function $f(\alpha Z)$. It is defined by:
\BE
f(Z\A )=\gamma_E + Re \Psi(1+iZ\A)= (Z\A )^2 \sum_{n=1}^{\infty}
\frac{1}{n(n^2+(Z\A)^2)},
\label{eq:fzabehei}
\EE
where $\gamma_E = 0.57721...$ is the Euler-Mascheroni constant \cite{Euler:2007} and
$\Psi$ the logarithmic derivative of the $\Gamma$-function. 
Note that in this way higher order effects are established without
making use of perturbative QED. In fact, Feynman graphs up to infinite order
are effectively summed up by the use of these Sommerfeld-Maue wave functions.

\begin{figure}
\centering
     \includegraphics[width=6cm]{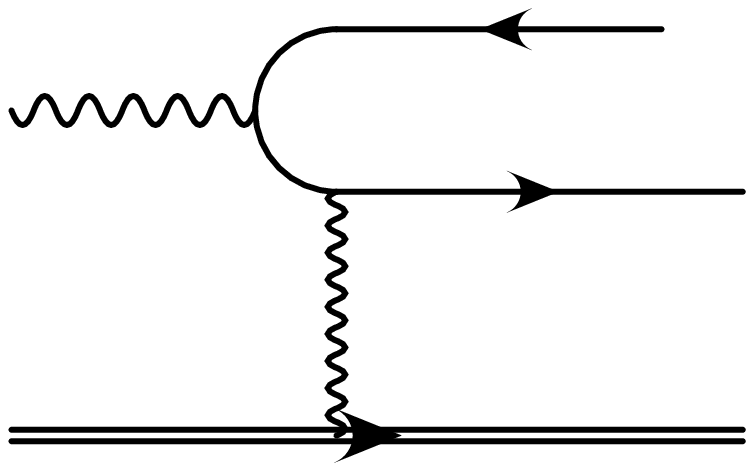}(a)~~
     \includegraphics[width=6cm]{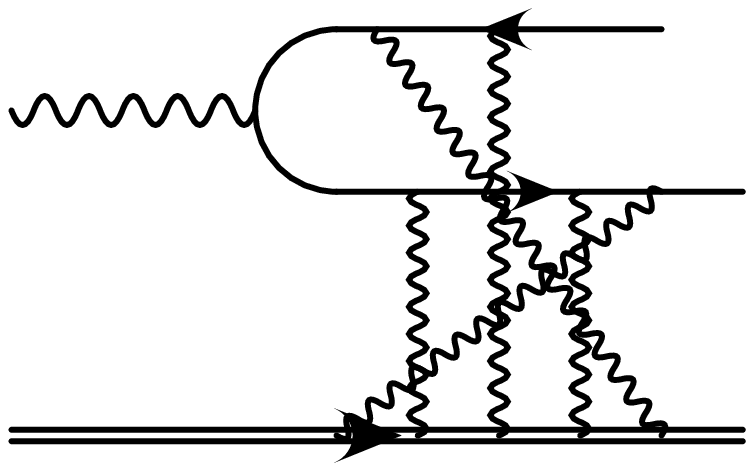}(b)
\caption{(a) Pair production in photon-nucleus scattering in lowest
  order of $Z \alpha$. (b) Higher order term. These terms are summed up in
the Bethe-Maximon theory using the Sommerfeld-Maue 
wave functions, or by summing the Feynman graphs according to
the approach of \protect\cite{Phys.Rev.D57.4025}.
}
\label{fig:betheheitler}
\end{figure}

It is very interesting to see that the same results can also be
derived directly (among other things) by the standard method of
summing up all Feynman graphs \cite{Phys.Rev.D57.4025}, see
Fig.~\ref{fig:betheheitler}. This establishes the equivalence 
between the standard higher order QED approach
and the calculations using Sommerfeld-Maue wave functions. 
This is very important for our problem below:
As discussed in the following chapters, there are again the 
Feynman diagram approaches to the nucleus-nucleus
problem, and the 'Sommerfeld-Maue type' of approaches.

In the following, we very briefly describe the work of Ivanov and Melnikov
\cite{Phys.Rev.D57.4025}.
A typical type of Feynman graphs to be calculated is shown in Fig.\ref{fig:betheheitler}, (which is
Fig.~1 of \cite{Phys.Rev.D57.4025}). For $N=1$ this corresponds to the 
Born approximation: the $\EPEM$ pair is produced by the 
collision of a real photon 
(which can also be virtual in \cite{Phys.Rev.D57.4025}) 
and a virtual photon coming from the nuclear Coulomb field.  

Many steps are needed to sum up the Feynman graphs
in the high energy limit. We sketch some of the important points:
The basic kinematics is $p_1+p_2= p_1^{'}+ q_1 + q_2$ and
it is convenient to regard the nucleus as a light particle with 
mass $m$ and charge $Z$. 
We assume $2p_1 \cdot p_2 \equiv s \gg m^2, Q^2$.  We denote $p_1 ^2=-Q^2$;
for a real photon we have $Q^2=0$. There are two important scales for the square
of the transverse part of the transferred momentum $q\equiv q_1 +q_2-p_1$:
${\vec q}\;^2 \sim m^2, Q^2$ and ${\vec q}\;^2 \sim m^4/s, Q^4/s$. The second 
low momentum transfer region
is only important for the case where one photon is exchanged, it
is responsible for the logarithmic rise of the cross section.
For the case of more than one photon exchange with the nucleus the
first region is the dominant one. 

For the high energy limit considered in \cite{Phys.Rev.D57.4025} 
the so-called Sudakov variables are used as the appropriate variables.
They are light-like vector $\tilde p_1$ and $\tilde p_2$, which are
pointing almost in the direction of $p_1$ and $p_2$. They are
given by
\BE
\tilde p_1 = p_1 + \frac{Q^2}{s} p_2, \qquad
\tilde p_2 = p_2 - \frac{m^2}{s} p_1.
\EE
These Sudakov variables can also be related to the so-called
light-cone variables. The use of these variables allows to express the
four-dimensional loop integrals occuring in the calculation of the
Feynman diagrams in terms of the transverse momentum only. The
transverse momentum integrals themselves are expressed in coordinate
(impact-parameter)space and the high energy amplitude
contains the so-called impact factors. These are scalar and vector 
structure quantities for the exchange of $N$-photons. Using recursion relations 
and other tricks to which we can only refer to the original paper 
\cite{Phys.Rev.D57.4025} the authors are able  to sum them.

The full result for the cross section for the pair production due to a 
virtual photon $\gamma ^*$ and where all the graphs
with $N=1$ to $N=\infty$ are summed is given in
Sec. IV B of   \cite{Phys.Rev.D57.4025}.
For a real incident photon the result was found to be identical to 
the one obtained previously by using the Sommerfeld-Maue wave functions.
The methods of \cite{Phys.Rev.D57.4025} can also be applied to 
the relativistic heavy ion case, which is the subject of the next section.

Just as in \cite{Phys.Rev.D57.4025}, we are especially interested in
the high energy limit, but it is also of  more general interest to study 
corrections in $\frac{1}{\omega}$. For this question we refer to Ref.
\cite{Phys.Rev.A69.022708}.

We note that the theoretical results 
in the high energy small scattering angle limit are 
symmetric with respect to the exchange of $e^+$ and $e^-$. 
It is certainly not the case for
bound-free pair production, a process to be discussed in Sec.~\ref{section:multred},
see \cite{PHRVA.A50.3980,Meier:2000ga}. 
We also mention without further discussion that another 
asymmetry was introduced by Brodsky and
Gillespie \cite{Phys.Rev.173.1011} in their second order Born treatment
of large angle $\EPEM$ pair production.

\section{Lepton pair production in relativistic heavy-ion collisions
using Feynman diagrams}
\label{section:fdia}

In this paragraph we want to briefly review the work on $\EPEM$ 
pair production in relativistic heavy ion collisions based 
on the summation of Feynman graphs in the high energy limit.
The problem is formulated in \cite{Bartos:2001jz}.
The process to be studied is 
\BE
A(p_1)+ B(p_2) \rightarrow e^+(q_+)+ e^-(q_-) + A(p_1^\prime)+B(p_2^\prime),
\EE
where the corresponding four-momenta are given in the brackets.

In the paper \cite{Bartos:2001jz} the first terms in the amplitude of 
$\EPEM$ pair production in the Coulomb field of two relativistic heavy ions
are calculated. The Sudakov technique is used which simplifies 
the calculations at high energies, see Sec.~\ref{section:behei}.
In lowest order in $\alpha$ (corresponding to class (i), see
Fig~\ref{fig:classesitoiv}) the classic result of Racah is obtained
\BE
\sigma=\frac{\alpha^4Z_1^2 Z_2^2}{\pi m^2}\frac{28}{27}(L^3-2.2L^2+ 3.84 L -1.636),
\label{eq:Racah}
\EE
where $L\equiv \ln(\gamma_1 \gamma_2)=\ln(\gamma^2)$ for symmetric collisions.
For details we refer to reference \cite{Bartos:2001jz}. Also further references
are given there, which actually give the steps needed to derive the
Racah formula by their methods.

Next they consider Coulomb corrections. The first correction
corresponds to the case where one nucleus emits one and the 
other nucleus emits two photons ($n=1, n^\prime=2$), corresponding to
the lowest order diagram of class (ii). This contributes
to the Coulomb corrections with the pair produced in an $C$-odd state.
Altogether six Feynman diagrams contribute to this amplitude.

The first $C$-even correction involves graphs where $n=1, n^\prime =3$
(belonging again to class (iii)) and $n=2, n^\prime=2$ (the lowest
order diagram contributing to class (iv)).
Whereas the former case proved to be tractable analytically,
obstacles were found  for the second case.
An explicit calculation of the $n=2$,$n'=2$ case
is reported in \cite{Bartos:2004ss}. 
Their result suggests that the familiar
eikonalization of Coulomb distortions breaks down for
the oppositely moving centers.

Lepton pair production in relativistic 
ion collisions to all orders in $Z\alpha$ with
logarithmic accuracy is studied in \cite{Gevorkyan:2003dp}. 
The amplitude is separated into the four terms
\BE
M = M_{(i)} + M_{(ii)} + M_{(iii)} + M_{(iv)}
\EE
corresponding to the four different classes defined in the introduction.
Different terms in the absolute square of the amplitudes
can then be classified according to their energy dependence,
i.e. by their dependence on different powers of the large logarithm 
$L=\ln(\gamma _1 \gamma _2)=\ln(\gamma^2)$. The absolute square of the 
Born amplitude $|M_{(i)}|^2$ leads to the famous rise of the cross
section with $L^3$. It should be noted that the Racah formula (see
Eq.~(\ref{eq:Racah})) contains also terms proportional to $L^2$ and $L$.
As discussed in \cite{Bartos:2001jz}   the next powers of $L$ come from the 
absolute square of $M_{(ii)}$
and $M_{(iii)}$ and their interference with the Born amplitude $M_{(i)}$.
The leading term of their result coincides with the one found by
other methods, which will be described below.
Their result for the Coulomb corrections in order $L^2$ is of the 
'Bethe-Maximon type':
\BE
\sigma_c= \frac{28 \alpha^4 Z_1^2 Z_2^2}{9 \pi m^2}(f(Z_1 \alpha) f(Z_2 \alpha
)L^2+ O(L)),
\EE
where $f(Z\alpha)$ was defined in Eq.~(\ref{eq:fzabehei}).

\begin{figure}
\centering
     \includegraphics[width=10cm]{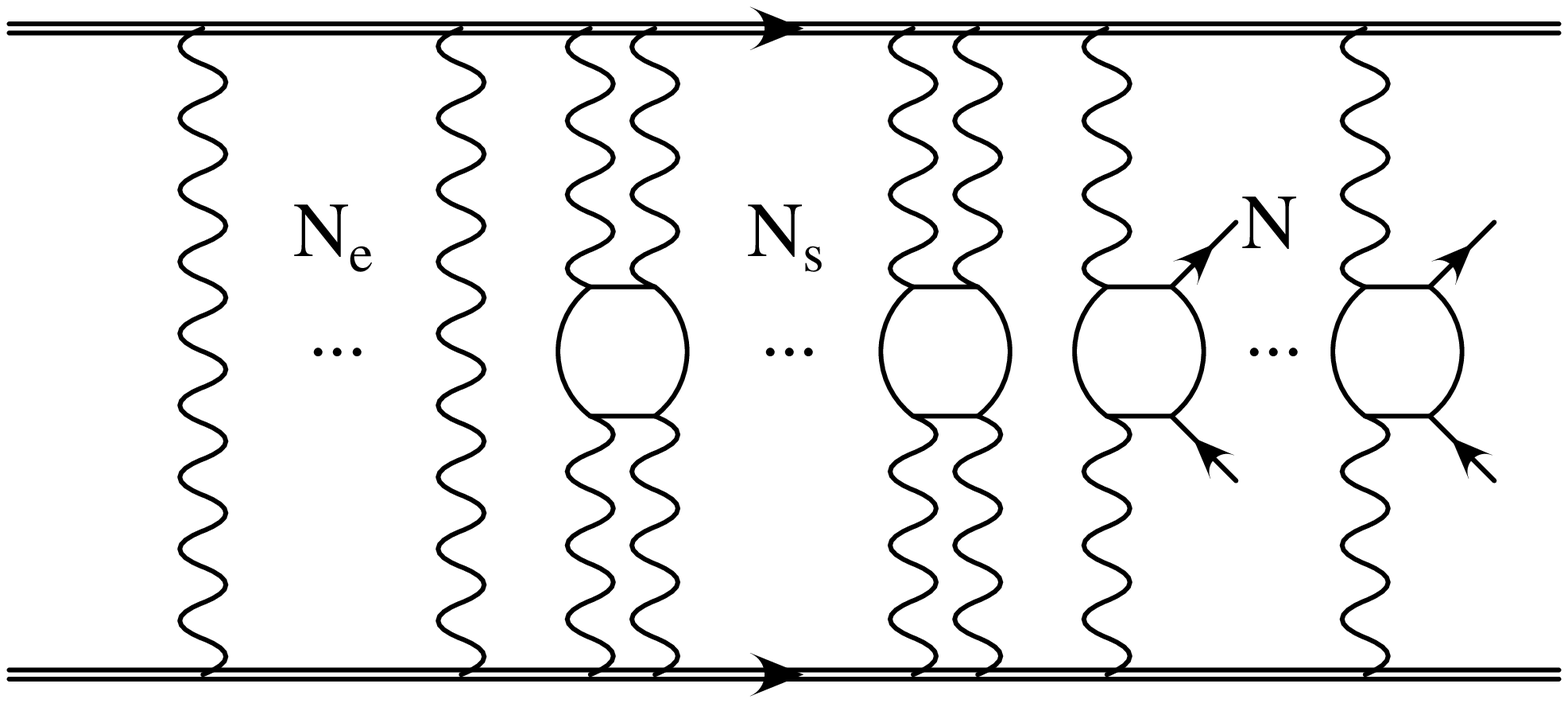}(a)\\
     \includegraphics[width=10cm]{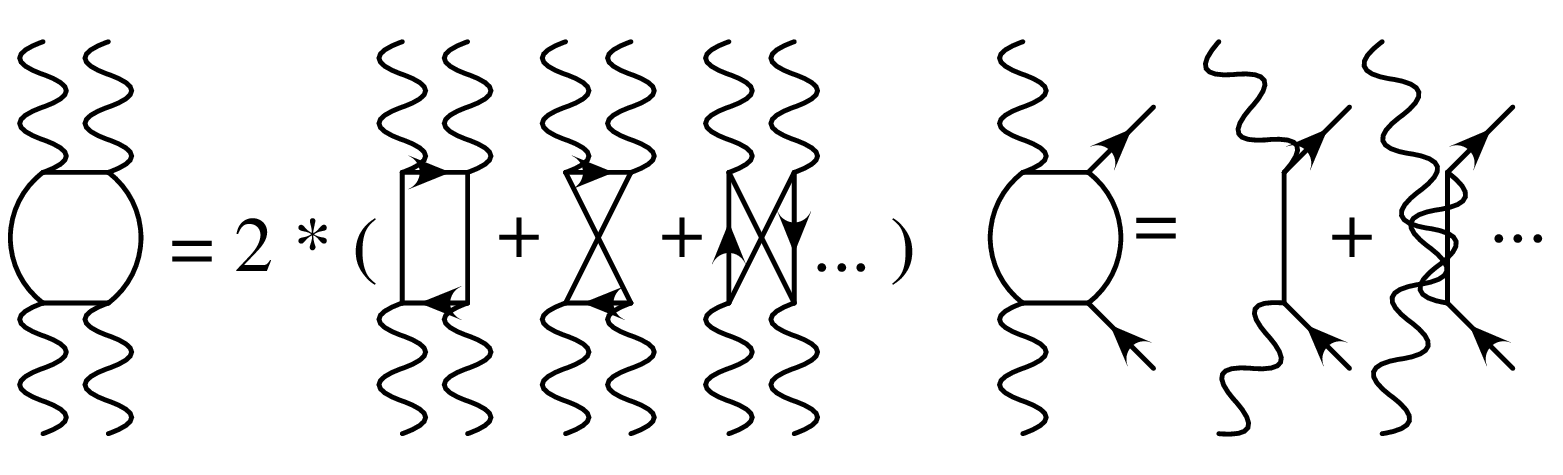}(b)
\caption{
 A general diagram for the process of $N$-pair production 
in ultraperipheral nucleus-nucleus collisions is shown. It contains 
$N_e$ photon exchanges between the nuclei, $N_s$ light-by-light scattering 
blocks and $N$ $\EPEM$ pair production diagrams. The symbols 
for light-by-light scattering blocks and pair production diagrams
are given in terms of the Feynman graphs in Fig. (b). 
}
\label{fig:bartoblocks}
\end{figure}

A direct manifestation of strong fields is multiple pair production. 
Early work on this subject started with the observation that 
the impact parameter dependent total pair production probability
calculated in lowest order perturbation theory is 
larger than one, i.e. $P^{(1)}(b)>1$.
More on multiple pair production will be given in 
Sec.~\ref{section:multred}.
The analysis of \cite{Bartos:2002ii}
fits in naturally within the approach to study Feynman graphs 
in the high energy limit.
The general Feynman diagram is shown in Fig.~\ref{fig:bartoblocks}.
It has $N_s$ 
light-by-light (LBL) blocks $N_e$ photon exchanges between the ions 
$Z_1$ and $Z_2$ (not present in an external field approach) and $N$ 
pair production processes. Coulomb corrections are neglected. They
assume  $Z_{1,2}\gg 1$, $ Z_1 \alpha \sim Z_2 \alpha <1$
and $Z_1 Z_2 \alpha >1$. The probability of $N$ pair production is
shown to obey a Poisson distribution.

In \cite{Gevorkyan:2005ys} a new approach to lepton pair production 
in the Coulomb field of two relativistic nuclei was developed. 
It was shown that for certain sums of finite terms
of the Watson series cancellations lead to infrared stability of the 
amplitude (i.e. the result does not depend on a screening radius 
of the Coulomb field).

\section{Exact solution of the one-particle Dirac equation 
in the ultrarelativistic 
limit and what one can conclude from that}
\label{section:mclerran}

An interesting topic is the solution of the (single-particle)
Dirac equation for two countermoving ions and how this relates
to the process of pair production. This was strongly triggered
by the observation, that in the limit of ultrarelativistic nuclei and
using an appropriate gauge, the expression for the electromagnetic 
interaction simplifies and the Dirac equation can be solved
analytically in a closed form \cite{Baltz:1991xx}.

Before we discuss this specific solution of the Dirac equation,
it is useful to remember that $\EPEM$ pair production is 
a problem in relativistic quantum field theory. It is inherently a
many-particle theory from the beginning, where particles can be 
created and destroyed. 
In contrast to this situation the familiar problem of ionization 
of an atom by a non-relativistic heavy projectile can indeed be 
dealt with by the time-dependent Schr\"odinger equation: the atomic 
electron (say in a hydrogen atom) is described by a Hamiltonian 
$H_0=T+ e^2/r$. The classical motion of the projectile
causes a time-dependent perturbation $V(t)$ and the problem is 
to solve the time-dependent Schr\"odinger equation for the wave
function $\Phi$ of the electron
\BE
 i \frac{\partial \Phi}{\partial t}=(H_0 + V(t) )\Phi(t).
\label{eq:schroedi}
\EE
This equation can be solved by a number of well-established methods
like perturbation theory, see e.g. \cite{AlderW66}. For special cases, 
like the harmonic oscillator, there are also full analytical solutions.
Alternatively, one can solve numerically the Schr\"odinger equation for
the wave function $\Phi(t)$ and project on specific asymptotic states
in order to determine the cross-section for transitions to
these final states.

%KH: mein Vorschlag
In the case of $\EPEM$ pair production Eq.~(\ref{eq:schroedi})
has a different meaning: $\Phi$ is now
a state vector, the term $V(t)=j_\mu A_\mu$ is the interaction of the
time-dependent (external)
field $A_\mu(t)$ with the electromagnetic current operator  
$j_\mu=\bar \Psi \gamma_\mu \Psi$. This current can be written in terms of
electron and positron  
creation and annihilation operators. One gets different terms 
describing electron and positron
scattering, as well as, $\EPEM$ pair production and annihilation.
This problem is solved by the Dyson expansion 
(see, e.g., \cite{Landau:1986aa} or other textbooks on Quantum field
theory). This approach then leads to amplitudes for $\EPEM$ pair 
production not only for single pair production, but for
$N=0,1,2,\ldots$ pair production. Multiple pair production will be discussed later
in Sec.~\ref{section:multred}. Here we want to discuss only the relation
between a solution of the Dirac equation and the pair production process.
That there is a problem with the interpretation of the 
one-body Dirac equation can already be seen in the discussion of the 
free Dirac propagator, see e.g. Chap.~2-5 of \cite{ItzyksonZ80}. 
The solution of the (free) Dirac equation
(with $V(t)=0$) can be written as
\BE
\Psi(t_2, \vec x_2)=\int d^3x_1 K(t_2,\vec x_2; t_1, \vec x_1)
\gamma_0 \Psi(t_1,\vec x_1),
\EE
where $K(x_2, x_1)$ is the retarded kernel. However,
as is remarked on  p.~90 of \cite{ItzyksonZ80}, the hole theory
suggests the introduction of a different propagator, 
the Feynman propagator. It appears in a natural way in the quantized 
field theory.
% For interacting particles the Dirac sea picture,
%that is (to quote S. Weinberg, who quotes J. Schwinger \cite{WeinbergI})
%``the picture of an infinite sea of negative energy electrons
%is now best regarded as a historical curiosity and forgotten''.

In our case here, that is, for the interaction with only an external field
we can find at least a
relation between the two pictures, which we quickly want to sketch here. For
further details we refer to \cite{Aste:2001te,Hencken:1994hp}.
Only for certain processes one can neglect the many-particle aspect.
This is for example the case for photoproduction, where the
interaction with only one photon and a static Coulomb field leads to
only one electron lifted from the negative sea to the positive continuum.

%KH: Mein Vorschlag
Baltz, Gelis, McLerran and Peshier
discuss the meaning of the exact solution of the Dirac equation
in Ref. \cite{nucl-th/0101024}.
They make a valid point: the interpretation of the solution of the  
one-body Dirac equation is not straightforward. They show that
the total cross-section obtained in this approach
does not correspond the exclusive cross-section to produce exactly
one pair, but to one where multiple pair production cross section are
also included (each weighted by the number of particles produced). 
This is rederived in a very straightforward manner in
\cite{Aste:2001te}.

The relation between the two approaches can be found in the following
way, see \cite{Aste:2001te} or \cite{Phys.Rev.76.749,BialynickiB75}
for details. The $S$-matrix of the Dirac equation
relates the annihilation operators for the electrons $b$ and 
creation operators for positrons $d$, respectively, 
in the initial ($i$) and final ($f$) state
in the following way:
\BE
\left(\begin{array}{c}b^f\\d^{f+}\end{array}\right)
= S \left(\begin{array}{c}b^i\\d^{i+}\end{array}\right) =
\left(\begin{array}{cc}S_{++}&S_{+-}\\S_{-+}&S_{--}\end{array}\right)
\left(\begin{array}{c}b^i\\d^{i+}\end{array}\right).
\label{eq:smatrix}
\EE
Here the indices $+$ and $-$ refer to the positive or negative energy,
that is $S_{-+}$ is the transition from a positive to a negative energy
state. It is this matrix $S$ that is calculated by solving the Dirac
equation. 

For the Feynman boundary condition one 
prescribes the initial electron and the final positron (following
the picture of Feynman that the positrons are electrons traveling back in
time) and solves the field equations to get the final electron
and initial positron states. For single pair production we have one positron
in the final state and calculate those processes leading to an
electron in the final state.
Therefore we need to rewrite the relation in Eq.~(\ref{eq:smatrix}) as
one between the $b^i$ and $d^{f+}$ on the one and the $b^f$ and
$d^{i+}$ on the other side. One gets 
%For this we multiply the lower one of the
%equations above by $s_{--}^{-1}$, the inverse of $s_{--}$ alone to get
%\BE
%d^{i+}=s_{--}^{-1} d^{f+} - s_{--}^{-1} s_{-+} b^i.
%\EE
%We use this to replace $d^{i+}$ in the upper part
\BE
b^f = (S_{++} - S_{-+} S_{--}^{-1} S_{-+}) b^i + S_{+-} S_{--}^{-1} d^{f+}.
\EE
The first term is of interest for electron scattering, whereas the
second term is the one we are interested
in. In addition we need to add the vacuum-vacuum transition amplitude
$C$, see Sec.~\ref{section:multred} below, to get for the matrix
element for the single pair production process
\BE
M_{kl} = \langle 0|b^f_k d^f_l |0\rangle = C (S_{+-} S_{--}^{-1})_{kl}.
\EE

Having a complete solution of the Dirac equation one can now in
principle invert $S_{--}$ (the amplitude for the scattering of positrons
to positrons) in order to get the correct single-pair production
amplitude.
%
%
%\begin{figure}
%\centering
%     \includegraphics[width=3cm]{figs/multiparticle.ps}
%\caption{One contribution to the pair production process, where
%  many-particle aspects come into play. This diagram was estimated to
%  contribute to about 5\% to the probability at small impact parameter.
%}
%\label{fig:multiparticle}
%\end{figure}

\begin{figure}
\centering
     \includegraphics[width=9cm]{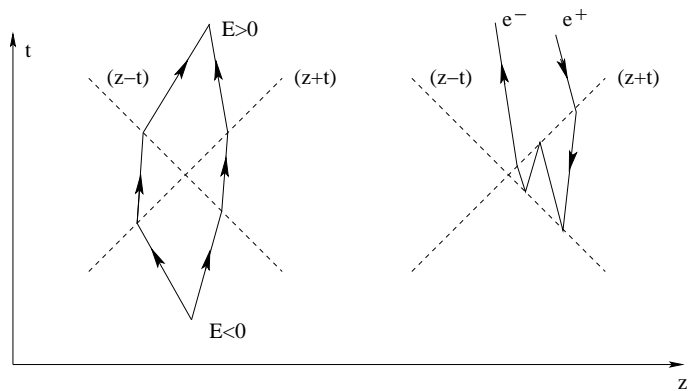}
\caption{
The special structure of the electromagnetic interaction can
be seen in this t-z plot. Due to the Lorentz contraction, the electromagnetic
fields are localized in two sheets corresponding to $z=\pm t$. In the 'retarded'
or 'Dirac sea' approach (left), an electron with negative energy comes
from $t=-\infty$, crosses the field of each ion only once before
leaving as an electron with positive energy.
In the 'Feynman' approach (right), the positron comes from $t=+\infty$ and can go
forward and backward in time interacting a number of times with the ions. 
}
\label{fig:lightcone}
\end{figure}

In the papers of Segev and Wells \cite{Phys.Rev.A57.1849,Phys.Rev.C59.2753}, 
Baltz and  McLerran \cite{Phys.Rev.C58.1679}
and Eichmann, Reinhardt, Schramm and Greiner \cite{Phys.Rev.A59.1223}
a remarkable analytic solution of the 
(one-body) Dirac equation is found. It was shown 
subsequently by Lee and Milstein how to regularize this expression and
to find a useful solution for the Coulomb correction problem
(see the next Sec.~\ref{section:serboetal}).
By a suitable gauge transformation \cite{Phys.Rev.C58.1679}
the potential created 
by the counter-moving ultrarelativistic ions can be written as
\BE
V(\rho,z,t)=-\theta (t-z) \vec \alpha \cdot \nabla \Lambda^-(\rho)
- \theta(t+z)\vec \alpha \cdot \nabla \Lambda^+(\rho),
\label{eq:Vsudden1}
\EE
where 
\BE
\Lambda^\pm(\rho)=-Z\alpha \ln\frac{(\vec \rho \pm \vec b/2)^2}{(b/2)^2}.
\label{eq:Vsudden2}
\EE
Except at $x^\pm \equiv \frac{1}{\sqrt 2}(t \pm z)$ the electron or positron
propagates as a free particle. As is remarked in \cite{Phys.Rev.C58.1679},
to construct the propagator for the electron, one needs to solve a
boundary value problem with free propagation everywhere except at the
surfaces of discontinuity at the light cone $x^\pm=0$, see Fig.~\ref{fig:lightcone}.
The final result is their Eq.~(51). We do not give it here but refer to the 
next section, where the work in \cite{Lee:1999ey} and~\cite{Lee:2001aa} 
is discussed. There one starts from their formula
Eq.~(\ref{eq:Vsudden1}), which is now 
contained in Eqs.~(\ref{eq:baslm})--(\ref{eq:chirho}) of Sec. \ref{section:serboetal}
and show how to regularize this result correctly.

\section{Coulomb corrections for the heavy ion case,
based on the Bethe-Maximon approach }
\label{section:serboetal}

In the last section  the exact solution of the 
one-particle Dirac equation in the high energy limit was discussed. 
Using the eikonal/sudden approximation at high
energies it was claimed in several publications that Coulomb
corrections are not present. This claim must be wrong, since one has
to reproduce the Bethe-Maximon results for
photoproduction, see Sec.~\ref{section:behei} in the limit where
the charge of one nucleus $Z_1$ is small, i.e. $Z_1 \alpha\ll1$.
In Sec.~\ref{section:fdia} we dealt with pair production
in the ``usual'' approach of QED, that is the perturbative calculation 
and summation of Feynman diagrams. We saw that in the papers which we 
briefly reviewed very powerful techniques could be used to simplify
the calculations at high energies. Also various terms could be summed
analytically. In the end an agreement with the Bethe-Maximon results
was found. Yet, people not so familiar with these techniques might
feel a little lost and other approaches (still for the high energy
case) might be preferable. This is the subject of the present section.

Before starting it may be mentioned that in Ref. 
\cite{PRPLC.163.299} it was already realized that there was a problem 
with higher order Coulomb corrections, see Chap.~7.3 there.
There are corrections of the Bethe-Maximon type (class (ii) and (iii)
in our notation) and corrections due to class (iv) (with $n, n^\prime>1$).
An 'interpolating' function $\bar{f}(Z_1, Z_2)$ was conjectured in 
\cite{PRPLC.163.299} (see Eq.~(7.3.7)), in order to take care of these
corrections but the question remained inconclusive.
The  convincing answer for the Coulomb corrections was first given 
in Ref. \cite{Ivanov:1998ru}.

\subsection{Coulomb corrections in the equivalent photon
approximation: the approach of Ivanov, Serbo and Schiller}
Let us now briefly review the approach of Ref. \cite{Ivanov:1998ru}.
These authors start with a classification of the pair production
amplitudes $M_{n n^\prime}$ according to the number of photon
lines attached to each ion, that is, the classification introduced
in Sec.~\ref{section:genintro}. The sum of the amplitudes 
is written as $ \sum_{n n^\prime} M_{n n^\prime} = M_{Born}
+ M_1 + \tilde{M_1} + M_2$, corresponding to 
$M_{(i)}
+ M_{(ii)} + M_{(iii)} + M_{(iv)}$ in our notation. 
The Born amplitude contains
the one-photon exchange with each nucleus (class (i)), 
the amplitudes $M_1$ and $\tilde{M_1}$ correspond to classes (ii) and
(iii), $M_2$ are the processes of class (iv). 
In their Eq.~(8) the total cross section is classified 
according to
\BE
\sigma= \sigma_{Born}+ \sigma_1 + \tilde{\sigma_1}
+ \sigma_2,
\label{eq:sigmadecomp}
\EE 
where
\BE
d \sigma_{Born} \sim |M_{Born}|^2 = |M_{(i)}|^2
\EE
and 
\BE
d \sigma_1 \sim 2Re(M_{Born} M_1^*) + |M_1|^2
 = 2 Re( M_{(i)} M_{(ii)}^*) + |M_{(ii)}|^2
\EE
etc.
Now the authors estimate the leading logarithms
$L\equiv \ln(\gamma_1 \gamma_2)=\ln(\gamma^2)$ 
(see Eq.~(\ref{eq:Racah}) in Sec.~\ref{section:fdia})
appearing in the cross sections $\sigma_i$. 
Due to the integration over the transferred momenta there are 
two large logarithms $L$ in $\sigma_{Born}\sim L^2$ and 
one large logarithm $L$ for $\sigma_1\sim L$ and $\tilde{\sigma_1}\sim
L$. There is no large logarithm for $\sigma_2 \sim L^0$.
(Eventually, there will be another logarithm due to the integration
over the energy of the $\EPEM$pair.)
Therefore this last (and most troublesome) term can be neglected, 
to a good accuracy, in the high energy limit when calculating the
cross section.
Please note that this argument is true for the cross section. As we
will discuss below it does not hold for 
impact parameter dependent probabilities, especially  
for small impact parameters $b$.

In order to study the higher order effects, the method
of {\em equivalent photons} is used. Thus the higher order effects 
are reduced to a study of the higher order effects in the 
photoproduction (by real photons). These corrections are well known,
see Sec.~\ref{section:behei}, 
and given in Eqs.~(10)--(13) of Ref. \cite{Ivanov:1998ru}. 
Their final result is the expression Eqs.~(24)--(27) for 
$\sigma_1$, and a corresponding one for 
$\tilde{\sigma_1}$. As discussed above, it is proportional to $L^2$.
Further, it is estimated
that $\sigma_2/\sigma_{Born} \sim (Z\alpha)^2/L^2$. Therefore this
last class of processes will give a contribution so small that one can
neglect this contribution for practical purposes.
In order to see these effects either a very good accuracy would be
needed, or one should find a way to increase their relative importance.
One possibility to enhance the importance of small impact parameters
is discussed in Sec.~\ref{section:compa}.
Their final result (Eq.~(27)) for the Coulomb correction is then
\BE
\sigma_{Coul} \approx \sigma_1 + \tilde{\sigma_1} = -\frac{56}{9}\sigma_0 f(Z) L^2,
\label{eq:cciss}
\EE
where $f$ is given in Eq.~(\ref{eq:fzabehei}) in Sec.~\ref{section:behei} and
$\sigma_0 =\frac{\alpha^4 Z_1^2 Z_2^2}{\pi m^2}$.

It is also noted that the large transverse momenta $\vec p^2 \gg m^2$
give a 
negligible contribution to the total cross section, but the
experiments of STAR at RHIC (see Sec.~\ref{section:compa} below) are just
in this region, since one needs the high transverse momentum 
to detect the electrons.
                           
\subsection{ How to regularize the exact solution of 
Baltz and McLerran and Segev and Wells: the approach of 
Lee and Milstein}

As shown above we know that there are indeed Coulomb corrections.
They are predominantly given by the Coulomb corrections 
which are present in the photonuclear case.

In order to see the Coulomb corrections in the
exact solution of the Dirac equation, we need to study in more detail
the regularization that is needed in the sudden limit. In this limit
all longitudinal momentum transfers are neglected. In
this case the cross section would diverge, as large impact parameters
contribute to arbitrary large values. Therefore a regularization of
the result is needed. This is done in a certain way, by adding a
longitudinal component to the final result in
\cite{Phys.Rev.C58.1679}. Another approach is to calculate only the
deviation of the full results from the Born result, as it is only the
Born result, which leads to a divergence. In
\cite{Lee:1999ey,Lee:2001aa} 
it is shown how to obtain reasonable results from the expression of
Segev and Wells \cite{segev:1998ur} and Baltz and McLerran
\cite{Phys.Rev.C58.1679}. The problem of the slowly decaying 
Coulomb potential is solved by introducing a screening of the 
Coulomb potential. Eventually the screening parameter is extended to
infinity. The procedure of \cite{Lee:2001aa} and \cite{Lee:1999ey} is 
a nice example of the power of analytic methods. In order to outline
their reasoning we would like to repeat some of the key steps of this work.
For details we have to refer to their work.

The authors of \cite{Lee:1999ey} start from 
an expression derived by Segev and Wells
\cite{segev:1998ur} and Baltz and  McLerran 
\cite{Phys.Rev.C58.1679} for 
the total pair production cross section, see their Eq.~(1)
\BE
d \sigma=\frac{m^2d^3pd^3q}{(2\pi)^6\epsilon_p \epsilon_q}
\int \frac{d^2k}{(2\pi)^2} |F_B(\vec k)|^2 |F_A(\vec q_\perp +
\vec p_\perp -\vec k)|^2|M(\vec k)|^2.
\label{eq:baslm}
\EE
Here $\vec k$ is a vector lying in the plane transverse to the momenta
of the ion(s).
\begin{figure}
\centering
     \includegraphics[width=6cm]{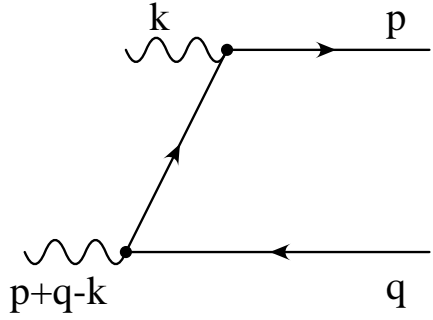}
\caption{One of the two lowest order diagrams for the pair production
  from two (virtual) photons with momenta $k$ and $p+q-k$.
}
\label{fig:ggee}
\end{figure}
The amplitude $M$ is given by 
\begin{eqnarray}
M(\vec k)=\bar u(p)\lbrack \frac{\vec \alpha \cdot (\vec k -\vec p_\perp)
+\gamma _0 m}{-p_+q_- - (\vec k -\vec p_\perp )^2-m^2} \gamma_-
\nonumber\\
+ \frac{-\vec \alpha \cdot ( \vec k -\vec q_\perp) + \gamma_0 m}
{-p_- q_+ - (\vec k -\vec q_\perp)^2-m^2}\gamma_+\rbrack u(-q),
\label{eq:matrixel}
\end{eqnarray}
where $\vec p$ and $\epsilon_p$ ($\vec q$ and $\epsilon_q$) are momentum
and energy of the electron (positron), $u(p)$ and $u(-q)$ are positive and negative energy Dirac spinors, $\vec \alpha=\gamma^0 \vec \gamma$, $\gamma_\pm\equiv
\gamma^0 \pm \gamma^z$, $\gamma^\mu$ are the Dirac matrices. We
use light-cone variables $p_\pm=\epsilon_p \pm p^z$ and $q_\pm = \epsilon_q
\pm q^z$. The function $F(\Delta)$ describes the interaction with the
Coulomb field of one of the ions. It is proportional to the 
electron eikonal scattering amplitude for one of the potentials $V(r)$:
\BE
F(\Delta)=\int d^2\rho \exp(-i \vec \rho \cdot \vec \Delta)
(\exp (-i\chi(\rho))-1),
\EE
where 
\BE
\chi(\rho)=\int_{-\infty}^{+\infty}
dz V(\rho,z).
\label{eq:chirho}
\EE
For an unscreened Coulomb potential $V(r)=-Z\alpha/r$ the 
integral for the Glauber phase $\chi(\rho)$ becomes divergent and 
a regularization is required. One can choose, e.g.,
$V=-Z \alpha \exp(-r/a)/r$, i.e., 
the Coulomb potential is smoothly cut off at a radius $a$.
Eventually, the radius $a$ will tend to $\infty$.
It is shown in \cite{Lee:1999ey,Lee:2001aa} that the result
is independent of the particular method used, as long as 
$V(r) \rightarrow -Z\alpha/r$ for $r \rightarrow 0$. 

Using Eq.~(9.6.23) of Ref. \cite{Abramowitz:1964aa}
and the substitution $dz=r dr /\sqrt{r^2-\rho^2}$
we find that $\chi(\rho)=\int dz V(b,z)=-2Z\alpha K_0(\rho/ a)$. 
Using the following well-known
representation of the Bessel function (see e.g.
\cite{Abramowitz:1964aa}) 
\BE
J_0(z)=\frac{1}{2\pi}\int_0^{2\pi}
e^{iz \cos \phi} d\phi,
\EE
one finds Eq.~(11) of \cite{Lee:2001aa}:
\BE
F(\Delta)=2\pi \int_0^\infty d\rho \rho J_0(\rho \Delta) (\exp{2iZ\alpha K_0(\rho/a)}-1) 
\label{eq:formf}
\EE
Let us now take the limit $a \rightarrow \infty$ for a fixed $\Delta \neq 0$.
For $\rho /a\ll 1$ we have $K_0 \sim \ln (\rho /a)$ and 
we get an integral of the type
\BE
F(\Delta) \approx 2 \pi \lim_{a \rightarrow \infty} \int_0^\infty d\rho \rho J_0(\rho \Delta)
(\exp{2i\ZA \ln(\rho /a)}-1).
\EE
This is a well-known generalization of the Bethe integral and one obtains 
\BE
F=i\pi \ZA \frac{\Gamma(1-i\ZA)}{\Gamma(1+i\ZA)}
\biggl(\frac{4}{\Delta^2}\biggr)^{1-i\ZA}.
\label{eq:famous}
\EE

Up to an overall constant phase, we see that the effect of the higher order
Coulomb interaction consists of a change of the photon propagator
$1/\Delta^2$ to $1/(\Delta^2)^{1-iZ\alpha}$). This result was already
found in \cite{Phys.Rev.C58.1679}. Taking the square of $F(\Delta)$
one gets
\BE
|F(\Delta)|^2 \sim \left| \frac{1}{(\Delta^2)^{1-i Z\alpha}} \right|^2 =
\frac{1}{\Delta^4},
\EE
independent of $Z\alpha$, which is the reason why the absence of
Coulomb corrections was found in their case. But one should keep in
mind, that $\Delta$ is the transverse part of the momentum transfer
from the ions only.
In \cite{Lee:1999ey,Lee:2001aa} a recipe is given that is used in
order to obtain the ``correct Coulomb correction''.
It is necessary to do the integration over $k$ first in 
Eq.~(\ref{eq:baslm}) using the function $F(k)$ with the regularized
phase and then to remove the
regularization only afterwards.

Before the Coulomb correction to electron positron pair
production itself is studied in refs. \cite{Lee:1999ey,Lee:2001aa}
these authors studied the integral
\BE
G= \int \frac{d^2k}{(2\pi)^2} k^2(|F(\vec k)|^2-|F^0(\vec k)|^2),
\label{eq:defg}
\EE 
where $F^0(\Delta)=-i\int d^2 \vec \rho \exp(-i\vec \Delta \cdot \vec \rho)
\chi(\rho)$ corresponds to the Born expression.

We integrate $F(\Delta)$ of Eq.~(\ref{eq:formf}) by parts over $\rho$ to
obtain:
\BE
F(\Delta)=\frac{2\pi i}{\Delta} \int_0^\infty d\rho \rho J_1(\Delta\rho)
\chi^\prime (\rho) \exp(-i\chi (\rho)).
\label{eq:byparts}
\EE
In order to obtain this result recursion relations for the Bessel
functions are used (see Eq.~(9.1.27) of \cite{Abramowitz:1964aa}).
The function $F^0(k)$ can be obtained from Eq.~(\ref{eq:byparts}) by
omitting the exponent in the integrand. Now the expression for $G$
reads 
\begin{eqnarray}
G&=&2 \pi \int_0^\infty dk k \int_0^\infty \int_0^\infty d \rho_1 d \rho_2 \rho_1 \rho_2
J_1(k\rho_1) J_1(k\rho_2) \chi^\prime (\rho_1) \chi^\prime(\rho_2) \nonumber\\
&&\times(\exp(-i\chi(\rho_1) + i\chi(\rho_2))-1).
\end{eqnarray}
A naive interchange of the order of integration would lead to $G=0$, since 
due to the orthogonality relation 
for the Bessel functions the integration over $k$ leads to a term proportional
to $\delta(\rho_1 - \rho_2)$.

Instead the authors of Ref.~\cite{Lee:1999ey} proceed in another way:
they integrate $k$ up to a finite fixed value $Q$, 
instead of $\infty$. 
According to Eq.~(11.3.30) of Ref.~\cite{Abramowitz:1964aa}
(for $\mu =\nu =0$) this gives their Eq.~(7).

The final result for $G$ is then found to be
\BE
G=-8\pi (Z\alpha)^2 f(Z\alpha),
\label{eq:valg}
\EE
where, again, we encounter the ubiquitous function 
$f(Z\alpha)$ of Eq.~(\ref{eq:fzabehei}) above. 

Now we proceed to the Coulomb correction itself. 
The Coulomb corrections are defined in the usual way discussed above
as the difference of the full result to the Born result. 
One obtains three terms, see Eqs.~(7) and~(8) of \cite{Lee:2001aa}.
In \cite{Lee:1999ey,Lee:2001aa} it is argued that the main
contribution to the  
integrals comes from the region of small $k$. Accordingly,
the matrix-element $M$ of Eq.~(\ref{eq:matrixel}), 
is expanded as 
\BE
M=k_i k^{\prime}_j M_{ij},
\EE
where $k^{\prime}= q+p -k$.
In \cite{Lee:2001aa} $|M_{ij}|^2$ is then calculated explicitly.
This decomposition into three terms corresponds to the decomposition
already discussed above in Eq.~(\ref{eq:sigmadecomp}):
\BE
\sigma= \sigma_{Born} + \sigma_A^c + \sigma_B^c + \sigma_{AB}^c
=\sigma_{Born}+ \sigma_1 + \tilde{\sigma_1}
+ \sigma_2.
\EE
We have $\sigma_{Born} \sim L_A L_B$, $\sigma_A^c
\sim G_A L_B$, $\sigma_B^c \sim L_A G_B$, and
$\sigma_{AB}^c \sim G_A G_B$. The quantities
$G_{A,B}$ are defined in Eq.~(\ref{eq:defg}) and their value given in
Eq.~(\ref{eq:valg}). The quantities 
$L_{A,B}$ are defined by
\BE
L_{A,B}=\int \frac{d^2k}{(2\pi)^2} k^2 |F^0_{A,B}|^2.
\EE

The necessary integrations are performed with logarithmic accuracy.
The $L^3$ behavior of the Born term is recovered. 
(Note that the full Racah result, see Eq.~(\ref{eq:Racah}) in 
Sec.~\ref{section:fdia}, contains terms proportional to $L^2$ and $L$ also).
The Coulomb corrections $\sigma_A^c$ and $\sigma_B^c$
agree with those of \cite{Ivanov:1998ru}, see Eq.~(\ref{eq:cciss}).
They show the typical $L^2=(\ln\gamma)^2$ behavior.
A new result is the last term, $\sigma^c_{AB}$, where the Coulomb correction
applies to both nuclei. It scales only with $L$. 
This shows, as discussed above, that it is smaller by a factor of
$L^2$ compared to the Born result and therefore is less important in the high energy limit.

In a recent numerical calculation by Baltz et al. these results
were checked by a calculation starting from the expression
Eq.~(\ref{eq:baslm}) directly and doing the integrations
numerically
\cite{Baltz:2004dz,Baltz:2003dy,Baltz:2005ay,Baltz:2006mz,Baltz:2007hw}.
We refer the reader to these papers for more details.

Calculations using the exact solution as given in
\cite{segev:1998ur,Phys.Rev.C58.1679} were already done for small
impact parameter. As discussed above the size relation between the 
different terms is only true for the cross section, but not for
the impact parameter dependent probability (see also
Sec.~\ref{section:multred} below). It was found that Coulomb corrections
are present at small impact parameter. This can reduce the probability
by up to 50\%. These results were confirmed by a calculation in
\cite{Lee:2006ze}.

\section{Higher Order Effects in Electron Pair Production:
Multiple Pair Production and Bound-Free Pair Production}
\label{section:multred}
\label{sec_leptons_mpairs}

In the previous sections we have mainly dealt with one special
kind of higher order effects, the so-called Coulomb corrections.
By definition they describe the effect of higher order Coulomb interactions 
on the one-pair production. They are well understood in terms of the 
Bethe-Maximon theory, or, equivalently, the approach of summing
Feynman diagrams \cite{Phys.Rev.D57.4025}. Higher order interactions 
can also lead to a second kind of processes: multiple pair production.
Multiple pair production is the most
important process of this type and will be discussed in this section.
There will also be Coulomb corrections to multiple pair production
and we will see that they can be incorporated into the existing framework.

In the high energy limit discussed in Sec.~\ref{section:serboetal} the 
oppositely charged electrons and positrons enter in a symmetric way. 
This is in contrast to the second order correction  for large
angle pair production of \cite{Phys.Rev.173.1011}, see the discussion in
Sec.~\ref{section:behei}. An obvious asymmetry also occurs when the 
positively charged nucleus interacts with the
electron to form a bound state ($K$-,$L$-,\ldots shell
capture). This will be dealt with in Sec.~\ref{ssec:boundfree}.
%Finally 
We conclude with some remarks regarding muon pair production.
In this case the finite size of the nucleus needs to be taken into account,
which then leads to differences in the importance of Coulomb
corrections and multiple pair production cross sections as compared to the $\EPEM$
case.

\subsection{Multiple pair production}
\label{ssec:multipair}

In heavy ion collisions with $\eta\equiv \frac{Z_1Z_2 e^2}{\hbar v}\gg
1$ a classical impact parameter $b$ can be defined.
At the high ion energies considered here the classical
path can be taken as a straight line, see e.g.
Chap.~2 of \cite{Baur:1998ay} or \cite{NUPHA.A729.787} for details. 
One can then calculate impact
parameter dependent probabilities  for pair production $P(b)$
in the semiclassical or Glauber approximations. The total
cross section is given by the integration of this probability
over all $b$.

There is a handy formula for the impact parameter dependent 
(single) pair production probability in $P^{(1)}$ in lowest order, as 
shown in \cite{Phys.Rev.A65.022101}. Quasireal photons from each
nucleus collide to produce an $\EPEM$-pair, the total cross section
is denoted by $\sigma_{\gamma \gamma}(s)$, where $s$ is the square of the 
c.m. energy. Using a simple version of the impact parameter dependent
equivalent photon spectrum 
(valid for $1/m \ll b\ll \gamma/m$)
and the relation $\int_{4m^2}^\infty \frac{ds}{s} 
\sigma_{\gamma \gamma}(s)
=\frac{14\pi}{9}\frac{\alpha^2}{m^2}$
one obtains \cite{Lee:1999ey,Lee:2001ea}
\begin{eqnarray}
P^{(1)}(b) &\approx& \frac{28}{9\pi^2} \frac{Z_1^2 Z_2^2 \alpha^4}{m^2 b^2}
\left[
2 \ln \gamma_{lab}^2 - 3 \ln (m b) \right]
\ln(m b) \quad\mbox{for $1\ll m b \ll \gamma_{lab}$} \nonumber\\
           &\approx& \frac{28}{9\pi^2} \frac{Z_1^2 Z_2^2 \alpha^4}{m^2 b^2}
\left[
\ln \frac{\gamma_{lab}^2}{m b}
\right]^2 \quad\mbox{for $\gamma_{lab}\ll m b \ll \gamma_{lab}^2$}.
\label{eq_pbapprox}
\end{eqnarray}

This equation displays nicely the dependence on the impact parameter
$b$ and the electron mass $m$. 
It also shows directly that for sufficiently large 
values of $Z_1, Z_2$ and $\gamma_{lab}$
the first order probability $P^{(1)}(b)$ exceeds one, i.e.,
unitarity is violated. We further note that this calculation 
underestimates the probability in the range of small impact parameters
$ b \sim \lambdabar = \frac{\hbar}{m c}$. The equivalent photon approximation used to derive
this approximation is not justified in this region. 
More exact numerical calculations need to be done. 

The impact parameter dependent probability $P^{(1)}(b)$ in lowest 
order was calculated numerically in \cite{Hencken:1994my,Guclu95}.
The quantity $P^{(1)}(b)$ found
there is larger than the one given by Eq.~(\ref{eq_pbapprox}), 
and values larger than one are possible at RHIC and LHC. As an example
$P^{(1)}(b=0)=3.9$ (1.6), 
and $P^{(1)}(\lambdabar)=1.5$ (0.6) \cite{Hencken:1994my} for LHC 
(RHIC) are found.

\begin{figure}
\centering
     \includegraphics[width=6cm]{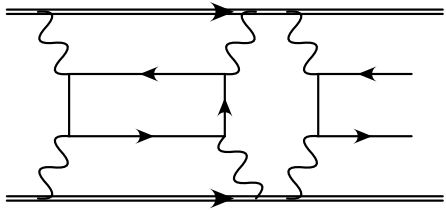}
\caption{
Third order contribution to the one-pair production.
}
\label{fig:thirdorder}
\end{figure}

The problem of the violation of the unitarity 
of $P(b)$, that is, that $P(b)$ 
will exceed unity for sufficiently large beam energies, 
was mentioned in  \cite{PRPLC.163.299}. It led 
to a series of studies starting around 1990. It was found that the 
production of multiple pairs in a single collision restores unitarity 
\cite{Baur:1990za}. $\EPEM$-pair production at high energies can be
studied in the sudden approximation, for not too large impact
parameters and invariant masses. I.e., we can neglect the
time-ordering operator in the Dyson-expansion of the $S$-matrix to get:
\BE
S= \exp(-i\int_{-\infty}^\infty V(t) dt).
\EE
One introduces the {\em pair} annihilation and creation operators $b_i$
and $b_i^\dagger$, where $i$ denotes collectively the state of the
pair (momenta and spin projection). To a good approximation the pairs
can be treated as ``quasibosons'' with the usual commutation relations:
\BE
[b_i,b_j^\dagger]=\delta_{ij}.
\EE
We can then write 
\BE
\int_{-\infty}^\infty V(t) dt= \sum_i u_i b_i^\dagger + u_i^* b_i,
\EE
where $u_i$ is a $c$-number which describes the probability amplitude
for the production of a pair in state $i$ in lowest order, see e.g.
Sect. 7.1 of \cite{PRPLC.163.299}.
In principle, the interaction $V$ also contains   
rescattering terms, i.e. terms, where the electron (or positron)
of a pair changes its momentum and spin. We neglect these terms. 
We use the Baker-Campbell-Hausdorf identity
\BE
e^{A+B}=e^A e^B e^{-\frac{1}{2}[A,B]},
\label{eq:quasiboson}
\EE
which is valid for two operators $A$ and $B$ for which the 
commutator is a $c$-number. With this identity one can directly obtain 
an expression for the production of multiple pairs with 
given quantum numbers, valid to all orders. In
Fig.~\ref{fig:thirdorder} we show a third order contribution 
to the one-pair production.
The other $\EPEM$ pair which is created  is annihilated (by a
``light-by-light scattering diagram'').
This is quite similar to the emission of multiple soft photons due to a 
classical current, see e.g. \cite{ItzyksonZ80}. One can say that a
coherent state \cite{Glauber:1963tx} of $\EPEM$ pairs is produced.

Similar approaches were followed in \cite{RhoadesBrownW91} and \cite{BestGS92}.
Another early work in this direction was given in \cite{Erlykin75a,Erlykin75b}.
All studies essentially make use of the ``quasiboson'' approximation, 
see Eq.~(\ref{eq:quasiboson}) above. It is then found that the probability
to produce $N$ pairs $P(N,b)$ is given by a Poisson distribution:
\begin{equation}
P(N,b) = \frac{P^{(1)}(b)^{N}}{N!} \exp\left[-P^{(1)}(b)\right],
\label{eq_poisson}
\end{equation}
where $P^{(1)}(b)$, the probability for pair production calculated in 
lowest order (which can become larger than one).
The quantity $P^{(1)}(b)$ acquires the meaning of the 
``average number of $e^+e^-$ pairs'' produced in a single ion collision: 
\begin{equation}
\left<N(b)\right> = \sum_N N P(N,b) = P^{(1)}(b).
\label{eq_averageN}
\end{equation}

\begin{figure}
\centering
     \includegraphics[width=6cm]{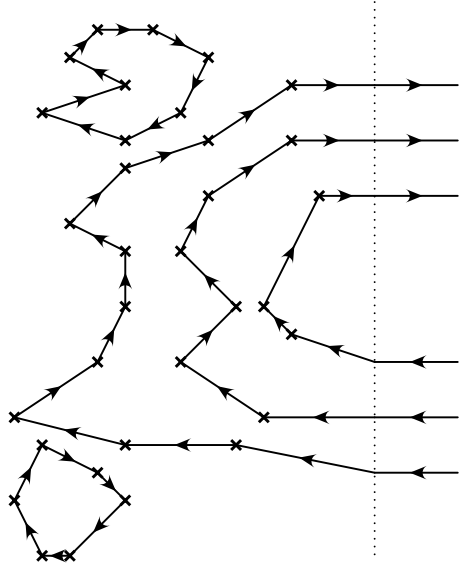}
\caption{
Graphical illustration of the $N$-pair production
process. The interaction with the external field is shown as crosses. The
production of a pair is described by a fermion line coming from and leaving
to the future, interacting an arbitrary number of times
with the external field. The vacuum-vacuum amplitude
$\left<0\right|S\left|0\right>$ corresponds to the sum of all closed
fermion loops.}
\label{fig:feynmanmulti}
\end{figure}

Another approach without the use of the quasiboson approximation was
done in \cite{Hencken:1994hp} and later also in
\cite{Baltz:2001dp}. There it was found that quite generally the
amplitude for $N$ pair production by an external field is given by
\begin{equation}
S_N = \left<0\right|S\left|0\right> \sum_\sigma \mbox{sgn}(\sigma)
s^{+-}_{k_1l_{\sigma(1)}} \cdots s^{+-}_{k_Nl_{\sigma(N)}},
\label{eq:mnpairs}
\end{equation}
where $k_i$,$l_i$ are the quantum numbers (momenta and spin projection)
of electron and positron, respectively, and $\sigma$ denotes a permutation of
$\{1,\cdots,N\}$, see also Fig.~\ref{fig:feynmanmulti}. The vacuum-to-vacuum amplitude is present in all QED
calculations, but is often dropped if it is of absolute value 1. Here
this amplitude
is $<1$ and is due exactly to the contribution of all
light-to-light diagrams, see Fig.~\ref{fig:thirdorder}. The amplitude $s^{+-}$
is the one for the production of a single pair, corresponding to 
a single fermion line. Neglecting the antisymmetrisation of the
amplitudes, one gets 
the $N$ pair amplitude as the product of the vacuum-vacuum amplitude
and the single-pair production amplitudes. From this one gets again a
Poisson distribution for the $N$-pair production probability. As
before the approximation introduced lead to an independence of each
pair production process from all others, which is the essential
ingredient needed to get the Poisson distribution. For further details
we refer to Chap.~6 of \cite{PRPLC.364.359}.

\begin{figure}[tbh]
     \includegraphics[width=12cm]{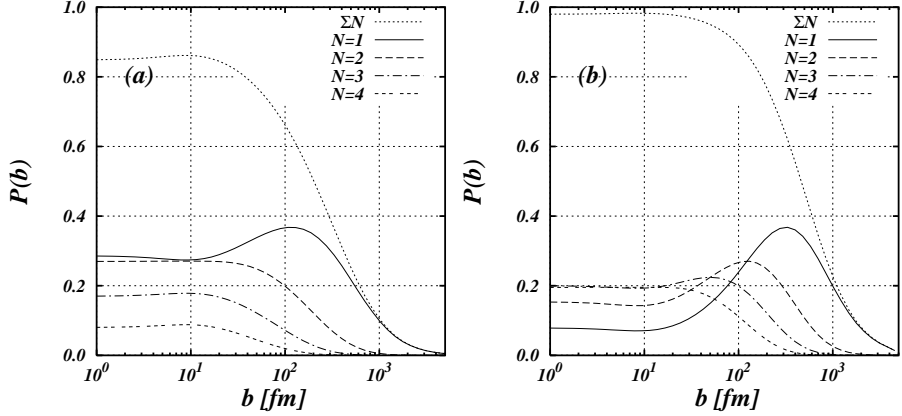}
\caption{
The impact parameter dependent probability to produce $N$
$\EPEM$-pairs ($N=1,2,3,4$) in one collision is shown for RHIC
($\G=100$, Au-Au) (a) and for the
 LHC ($\G=2950$,Pb-Pb) (b). Also shown is the total probability to 
produce at least one $\EPEM$-pair $\sum_{N=1}^\infty P(N,b) =  1-P(0,b)$. 
One sees that at small impact parameters multiple pair production 
dominates over single pair production.
}
\label{fig:pbee1}
\end{figure}

\begin{figure}
\centering
     \includegraphics[width=6cm]{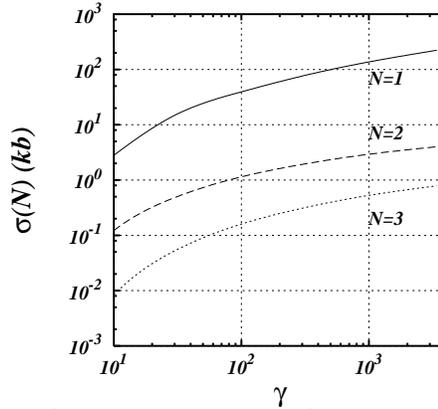}
\caption{Total cross section for the production of one, two and three
  pairs in PbPb collisions is shown  as a function of the ion energy.
}
\label{fig:pbee2}
\end{figure}

With the impact parameter dependent probability $P^{(1)}(b)$
calculated, e.g., in  \cite{Hencken:1994my,Guclu95} 
one can use Eq.~(\ref{eq_poisson}) to obtain the 
probabilities for $N$-pair production $P(N,b)$, see Fig.~\ref{fig:pbee1}. 
One can see that for impact parameters $b \approx 2R$ up to about $\lambdabar$
on the average 3--4 pairs will be produced in Pb-Pb collisions at the LHC.
This means that each photon-photon event --- especially those with high 
invariant mass, which occur predominantly at impact parameters close to 
$b \gtrsim 2 R$ --- is accompanied by the production of several (low-energy)
$\EPEM$ pairs (most of them however will remain unobserved experimentally).
Integrating over the impact parameter the total multiple pair production cross
section was given in \cite{Alscher:1996gn}, see Fig.~\ref{fig:pbee2}.

The Poisson distribution is also valid if, e.g., Coulomb corrections
are included in the calculation of the single pair production.
As seen in Eq.~(\ref{eq:mnpairs}) they need to be taken into account 
for a single lepton line to get a $P^1(b)$ including Coulomb
corrections. This $P^1(b)$ was calculated
for the high energy limit, as discussed in Sec.~\ref{section:serboetal}, in
\cite{Hencken:1998hf,Baltz:2006mz,Lee:2006ze}. It was found that the Coulomb
correction reduce $P^1(b)$ up to 50\%. 
The single pair production probability $P(b)$ falls off essentially 
as $1/b^2$. The total cross section 
is therefore dominated by large impact parameters and not very
sensitive to the 
multiple pair production effects at small $b$. In \cite{Lee:2001ea} 
the reduction of the exclusive one-pair production cross section $\sigma(N=1)$
was estimated to be -6.4\% for RHIC and -4.7\% for LHC, using the numerical
results of \cite{Hencken:1994hp,Alscher:1996gn}.

Multiple pair production cross sections are dominated by
impact parameters around $b\approx\lambdabar$.
It would be of interest to establish this effect experimentally.
In addition, Coulomb corrections are enhanced  for 
these close collisions.
At the SPS (CERN) the effect was looked for, see
\cite{VaneDDD97}, but only an upper bound could be given, which is
still above the theoretical prediction. The possibilities at the LHC
will be briefly discussed in \ref{section:compa}.

\subsection{Bound-free pair production}
\label{ssec:boundfree}

\begin{figure}
\centering
     \includegraphics[width=6cm]{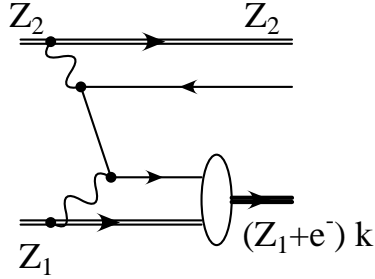}
\caption{Bound free pair production is the pair production process,
where the electron is produced into a bound state of one of the ions.
}
\label{fig:bfree}
\end{figure}

Bound-free pair production is the pair production process, where the
electron is created in a  bound state with  one of the
ions, say $Z_1$ (see Fig.~\ref{fig:bfree})
\BE
Z_1+ Z_2 \rightarrow (Z_1 + e^-)_{K,L,...} + e^+ + Z_2.
\EE

Since our previous review \cite{PRPLC.364.359} essentially no further
calculations of bound-free pair production were made.
The most careful state-of-the-art calculations were done
in \cite{Meier:2000ga}. It is a Born-approximation calculation, using Dirac
wave functions for the electrons and positrons in the field of a point nucleus
with charge $Z_1$. In this approximation the cross-section scales
with $Z_2^2$. At the high energies relevant for RHIC and LHC 
this is a good approximation, see also Sec.~\ref{section:compa}. Higher
order processes of the class (iv) would lead to a deviation from
the $Z_2^2$ scaling. They have been calculated using the ultrarelativistic
limit of the Dirac equation in \cite{Baltz:1996hr}. 
%But
Only a small reduction of the order of 1\% was found.

In an ``Einstein special issue''  electron-positron pair production
in relativistic heavy ion-atom collisions was reviewed
in \cite{Eichler:2005xn} in a general way, not restricted to the 
high energy limit. In this paper a discussion of the bound-free
pair production mechanisms is also given. While at the 
high energies the ``excitation type'' mechanism
studied in \cite{Meier:2000ga} and in  other
papers dominates, 
the ``transfer type'' mechanism is also important 
at lower beam energies. ``Excitation'' here means, that the
 electron and the positron are best described by wave functions in the
 target Coulomb field, ``transfer'' means that the positron is in the
 Coulomb field of the target, whereas the electron is ``transferred''
 to the Coulomb field of the projectile.  For the lower beam energies it becomes 
more proper to treat pair production by a two-center calculation.
At high energies, 'class (iv)' is relatively 
unimportant, see above. On the other hand, at lower energies this type
of process becomes increasingly important and causes further 
calculational complexities.

The cross-section for bound-free pair production at high energies rises 
only with $\ln \gamma$, i.e.,
\BE
\sigma=A \ln \gamma + B.
\EE
It is only a small fraction of the total cross section of
free pair production, which rises with $(\ln \gamma)^3$.
The parameters  $A$ and $B$ have been given for various systems in
\cite{Meier:2000ga}. Nevertheless, this process is 
an important one for the operation of the Pb-Pb beam at LHC:
the ions which have captured an electron from the vacuum
have a charge-to-mass-ratio different from the beam particles
and they will hit the wall at a rather well defined 'hot spot'.
This is discussed further in Sec.~\ref{section:compa} below.

\subsection{Higher order corrections to muon pair production}
\label{ssec:muonpairs}

In this review we focus on the production of electron-positron
pairs. Rather similar phenomena will appear in muon pair
production, 
scaled with the corresponding mass. The larger muon
mass has an effect on the different processes:
in \cite{Phys.Rev.D57.4025} it is remarked that the nucleus 
is an extended object with the inverse radius 
\BE
\Lambda = \frac{1}{R} \sim 30 MeV.
\EE
At such scales the electric field differs substantially from the pure
Coulomb behavior. This is especially of interest for muon pair
production, where we have the inequality
\BE
m \ll \Lambda \ll m_\mu.
\EE
This means that a simple scaling of the cross section by
$m^2/m_\mu^2$ is not valid. It is shown in \cite{Phys.Rev.D57.4025} 
that there is a suppression factor of $\Lambda /m_\mu$
for additional photon exchanges. This is because the form factor cuts out 
the short wavelength photons with $k>\Lambda$ and the remaining 
long-wavelength photons cannot resolve the dipole formed by the lepton pair
which is of extension $\sim 1/m_\mu$. This leads to a reduction of the
Coulomb effects for the muon case compared to the electron case. On
the other hand unitarity corrections, due to the additional production
of electron-positron pairs together with the muon-pair, is
enhanced. This can be understood as the impact parameter range in the
muon pair case is much smaller than in the electron case and the
probability to produce (additional) electron-positron pairs is then
large. For further discussion we refer to \cite{Phys.Rev.D57.4025}.

\section{Transition from the adiabatic to the sudden regime}
\label{chap:fastslow}
\label{section:fastslow}

The physics of $\EPEM$ pair production in fast ($v\approx c$) and slow
($v \approx 0$) heavy ion collisions ($Z_{\rm united \; \rm atom}>137(173)$)
is very different: 
for the slow collisions there is spontaneous pair production in supercritical fields,
a non-perturbative effect. For fast collisions we can apply perturbation theory. 
Two-photon production dominates, as was discussed above.
Let us first  mention the corresponding situation
of atomic ionization and electron-positron-pair production 
in a time-varying spatially-constant electric field of the form
\BE
\vec E(t)= \vec E_0 \cos\omega t.
\EE

The problem of ionization of atoms in a time varying
electric field is very similar to the problem of $\EPEM$ pair
production, which may be viewed as ionization 
of the negative energy Dirac sea, see Sec.~\ref{section:mclerran}. 
For ionization there is a similar transition from a slowly varying 
field to the high frequency case. The ionization of an 
H-atom in a time varying electric field is a textbook problem, 
see, e.g., p.~739ff, problem XVII in \cite{Messiah:1964aa}.
One limit --- large frequency $\omega$ and a
small field strength $E_0$ --- is the usual photo-effect.  The other limit is 
ionization in a static electric field. For a sufficiently strong
static ($\omega=0$) electric field, of the order of $E_a=\frac{m^2
  e^5}{\hbar^4} \equiv E_c \alpha ^3$, the electron 
disappears classically  into the continuum; for lower field
strength it can tunnel through the barrier, see, e.g., \cite{BetheS77}.

The critical atomic field strength $E_a$ is given by the condition
$e E_a a_B=\rm Ry$, 
i.e., the work done in the field over the typical length scale (the Bohr
radius $a_B$) is then equal to 
the typical energy scale (the ionization energy ${\mathrm Ry}=e^2/(2a_B)$).
(To obtain the critical field strength for the 
Schwinger case, discussed in Sec.~\ref{section:genintro}, one needs to
replace the Bohr radius by the 
Compton wavelength of the electron $\lambdabar=\hbar/(m c)$
and the Rydberg energy by the electron rest mass $m c^2$.) 
The field strength occurring in current lasers,
see, e.g., \cite{brabec:2000aa}
is strong enough to lead to such ionization processes:  in Chap.~VI 
of \cite{brabec:2000aa} the optical field ionization of atoms is
reviewed. The important parameter which divides the strong field 
and the weak field limits is the Keldysh parameter $\gamma_K$ 
(see Eq.~(21) of \cite{brabec:2000aa}). For
\BE
1/\gamma_K \equiv \frac{e E_0 a_B}{\hbar \omega}\ll 1
\EE
we have the perturbative limit, for 
\BE
1/\gamma_K \gg 1
\EE 
we have the strong field regime.
We do not go further into this but turn to $\EPEM$ pair production case,
which was studied in \cite{PhysRevD.2.1191}.

The adiabaticity parameter (see also
\cite{Ringwald:2003er,Ringwald:2001ib})
is given by Eq.~(47) of \cite{PhysRevD.2.1191}:
\BE
\gamma_{BI}=\frac{m c \omega}{ e E}.
\EE
It corresponds to the Keldysh parameter $\gamma_K$
of the atomic physics case. The Compton wave length
of the electron $\lambdabar$, now replaces the Bohr radius $a_B$.
Their final formula for the pair-production probability 
per unit time and volume is
\BE
w=\frac{\alpha E^2}{2\pi (g(\gamma_{BI})+1/2 g^\prime (\gamma_{BI}))}
\exp(-\frac{\pi m^2}{eE}g(\gamma_{BI})).
\EE
The function $g(\gamma_{BI})$ is given by Eq.~(45) of \cite{Phys.Rev.D2.1191},
it interpolates between the two extreme situations.
In the limit $\gamma_{BI} \ll 1$ it is given by
\BE
g(\gamma_{BI})=1-\frac{1}{8}\gamma_{BI}^2+ O(\gamma_{BI}^4)
\EE
and for $\gamma_{BI} \gg 1$ they find
\BE
g(\gamma_{BI})=\frac{4}{\pi \gamma_{BI}} \ln (2\gamma_{BI}) + O(1/\gamma_{BI}).
\EE
The parameter $\gamma_{BI}$ describes the transition from the 
high-field low-frequency limit ($\gamma_{BI}\ll1$) to the 
low-field perturbative regime $\gamma_{BI}\gg1$). 
In the $\gamma_{BI}\ll1$ limit, one obtains the Schwinger formula, see Sec.~\ref{section:genintro}.
For $\gamma_{BI}\gg1$  the perturbative result (see also Eq.~(11) of
\cite{Ringwald:2003er,Ringwald:2001ib}) is found:
\BE 
\frac{d^4n}{d^3x dt} \sim \frac{c}{4 \pi^3\lambda_e^4}
\biggl(\frac{e^2 E}{4\hbar \omega}\biggr)^{\frac{4m c^2}
{\hbar \omega}}.
\EE
It corresponds to $n$-th order perturbation theory, where
$n$ is the minimum number of quanta required to create an 
$\EPEM$ pair ($n \geq \frac{2m c^2}{\hbar \omega} \gg 1$).

For the heavy ion scattering the space-time dependence of the 
electric fields is much more complicated and one cannot obtain
such instructive analytic results.
Let us roughly  estimate a parameter $\gamma_{HI}$ for heavy ion collision 
corresponding to $\gamma_{BI}$:
we put $\omega_{max} \sim 1/\Delta t= \frac{\gamma v}{b}$. 
With the maximum field strength $E \sim \frac{Ze\gamma}{b^2}$,
see Eq.~(\ref{eq:maxfield}) in Sec.~\ref{section:genintro}, we obtain:
\BE
\gamma_{HI}= \frac{m c b v}{Z e^2}=\frac{p_{el}}{\Delta p},
\EE
where $p_{el}=mc$ is a typical electron momentum scale and
$\Delta p=\frac{2Ze^2}{bv}$ corresponds to the momentum transfer in a Coulomb
collision(see Eq.~(\ref{eq:deltat}) in Sec.~\ref{section:genintro}).
For a typical impact parameter $b$ one can take the 
Compton wave length of the electron, i.e., $b\approx\hbar/(m c)$ . So we have 
\BE  
\gamma_{HI}= v/(Z \alpha).
\EE
In the relativistic case, $v \sim c$, this  is $\ge 1$,  
i.e., we are in the low-field high-frequency limit.
For $v\ll1$ we are in the strong-field non-perturbative limit.

Our conclusion from this chapter is the following:
The slow collisions, with their static overcritical
fields are in a different regime of the adiabaticity parameter
as compared to the fast (ultra-) relativistic heavy ion collisions:
the perturbative approach, (sometimes up to spectacularly high orders,
see \cite{NUPHA.A729.787}) is well justified.
The crossover between these two regimes can then be expected to occur
in the region where the electron velocity is of the order of the velocity
of the ions, i.e. $v\approx Z\alpha$,
which corresponds to $\gamma\approx 1\ldots2$.

There is strong support for a future x-ray-free-electron 
laser at Desy (Hamburg), see \cite{Merkel:2006aa}
and one can expect exciting new results.
The possibilities of optics in the relativistic regime, which are opened
up by ultraintense laser pulses, are described in \cite{mourou:309}.
We mention the talk by A. Ringwald  
\cite{Ringwald:2004aa} or \cite{Ringwald:2003er}
and the theoretical calculations of pair creation in the fields of an
X-ray free electron laser \cite{Alkofer:2001ik}.
On the other hand, from what we said above, we have perturbative 
physics from the heavy ion collisions:
the fields get very strong, with increasing ion energy, but, at the same time,
the interaction time decreases in the same way.

\section{Comparison to experiment and outlook to 
LHC}
\label{section:compa}

Since our last review of this subject in \cite{PRPLC.364.359}
the heavy ion collider RHIC has come into operation. 
Both the STAR and the PHENIX collaborations
have published measurements on $\EPEM$ pair production. In the case of
STAR the data was taken for events accompanied by nuclear breakup. In
this way one is able to measure pair production at small impact
parameters. Due to the high Lorentz factor $\gamma$, the electromagnetic
 fields are stronger than at SPS or AGS energies. For a discussion of 
experiments at these accelerators we refer to previous reviews
\cite{PRPLC.364.359,Eichler90,Baur:1998ay}.
As explained in Sec.~\ref{section:fastslow} we are now in the 
perturbative limit. Also multiple pairs and bound free pair production
processes can be measured in principle. It is the purpose of this
section to give a short overview of these results and to compare them
with the theoretical analysis, especially to the one done in
\cite{Hencken:2004td}. Bound-free pair production was studied at RHIC
as well. In the end we give a short outlook to the possibilities at
the LHC.

\subsection{$\EPEM$ pair production in ultraperipheral
collisions at RHIC}

\begin{figure}
\centering
     \includegraphics[width=6cm]{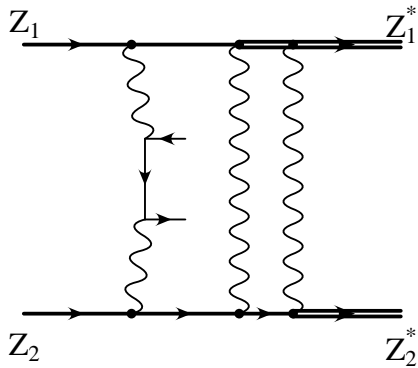}
\caption{At STAR pair production in connection with the
  electromagnetic excitation of one or both ions was measured.
}
\label{fig:pairplusgdr}
\end{figure}

The STAR collaboration has published data on $\EPEM$ pair production 
accompanied by nuclear breakup in ultraperipheral Au-Au collisions 
at a center of mass energy of 200 GeV per nucleon
\cite{Adams:2004rz}. This process is illustrated in Fig.~\ref{fig:pairplusgdr}.
The nuclear breakup of the gold ions is
predominantly due to the electromagnetic excitation of the 
giant dipole resonance (GDR), 
see \cite{PRPLC.364.359,Baur:1998ay}. 
The breakup neutrons are detected in the ZDC (Zero Degree Calorimeter).
It can
be shown, that the amplitudes for these processes factorize
\cite{NUPHA.A729.787} and that the cross section can be calculated by
integrating the product of the probabilities over all impact parameters
\BE
\frac{d^6\sigma_{e^+e^-,GDR(n)}}{d^3p_+d^3p_-} =
2 \pi \int_{b_{min}}^{\infty} b db (P_{GDR}(b))^n \frac{d^6P(b)}{d^3p_+ d^3p_-},
\EE
where $P_{GDR}(b)$ denotes the probability of exciting one
ion and $(P_{GDR}(b))^2$ is the probability to excite both ions. 
In this way pair production at small impact
parameter can be selected. The average impact parameter is given
approximately by \cite{NUPHA.A729.787}
\begin{equation}
  \overline b = \frac{\int d^2b\  b P(b)}{\int d^2b P(b)}
\approx
\frac{8R_a}{3} \approx 19\mbox{~fm},
\end{equation}

\begin{figure}
\centering
     \includegraphics[width=9cm]{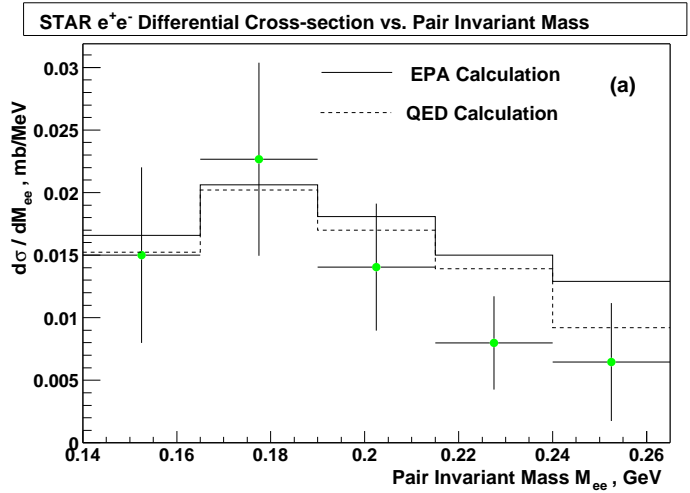}
\caption{Comparison of the distribution of the pair mass for the STAR
  experiment to the theoretical QED calculations of
  \protect\cite{Hencken:2004td} and to a calculation using the
  equivalent photon approximation (EPA). Taken from \protect\cite{Adams:2004rz}
}
\label{fig:stare}
\end{figure}

\begin{figure}
\centering
     \includegraphics[width=9cm]{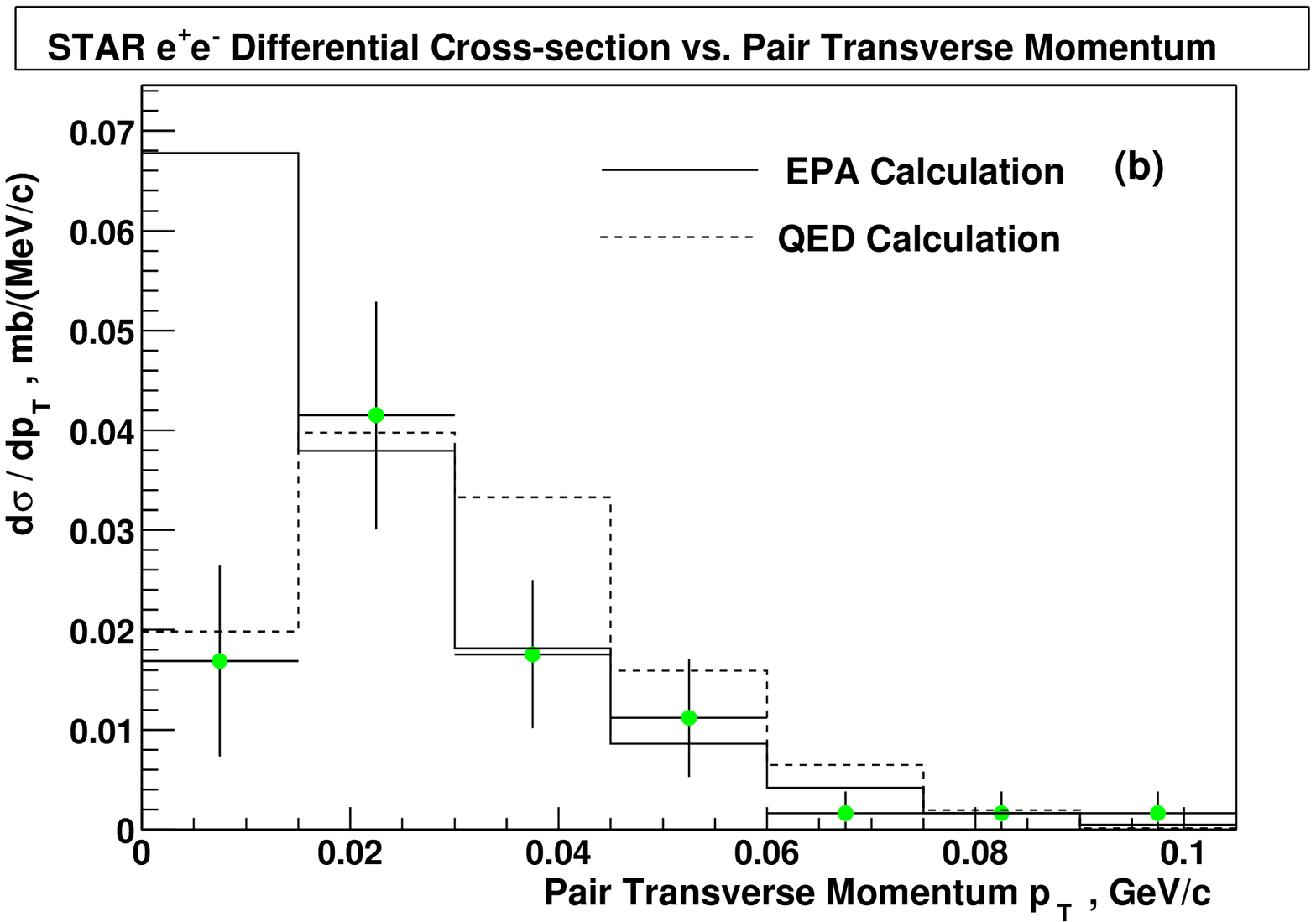}
\caption{Comparison of the transverse momentum distribution of the
  electron and the positron for the STAR experiment to the QED and the 
EPA calculation. Same as in Fig.~\protect\ref{fig:stare}. Taken from \protect\cite{Adams:2004rz}
}
\label{fig:starpt}
\end{figure}

\begin{figure}
\centering
     \includegraphics[width=9cm]{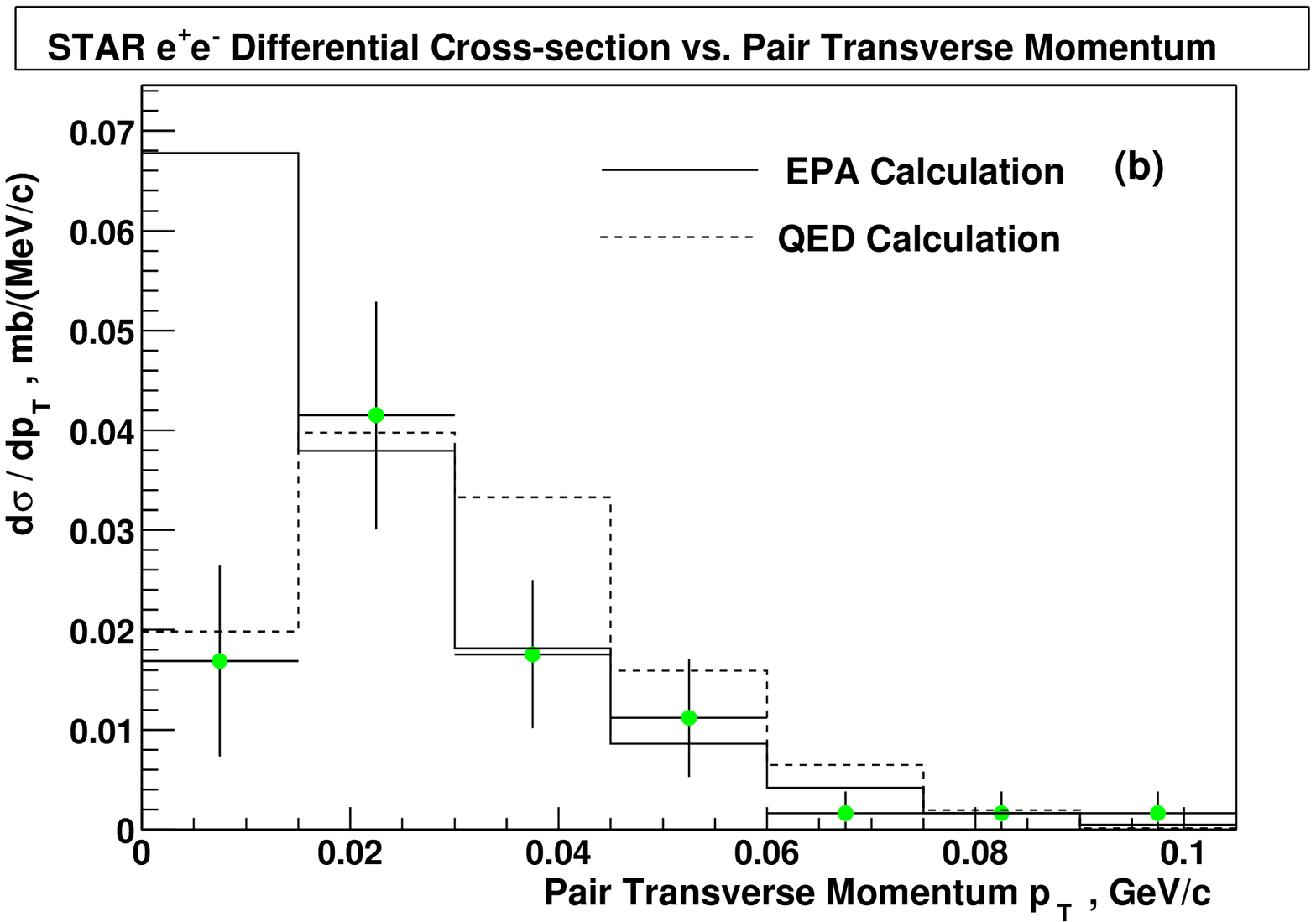}
\caption{Comparison of the transverse momentum distribution of the
pair for the STAR experiment to the QED and the
EPA calculation.  Same as in Fig.~\protect\ref{fig:stare}. 
Taken from \protect\cite{Adams:2004rz}.
}
\label{fig:starpp}
\end{figure}

The impact parameter dependent probability was calculated in lowest
order in \cite{Hencken:2004td} based on the method developed in
\cite{Hencken:1994my,Hencken:1993cf,Alscher:1996gn}. 
Only electrons and positrons with 
transverse momenta of $p_t> 60MeV/c$
with rapidity  $|y|<1.15$ could be detected. This limited
detector acceptance is taken
into account in the calculation. Within the experimental accuracy a
good agreement was found between the experimental and theoretical
results not only for the total cross section but also for the
distribution in energy and transverse momentum of electron and
positron, see Figs.~\ref{fig:stare}, \ref{fig:starpt}
and~\ref{fig:starpp} 
and in reference \cite{Adams:2004rz,Morozov:2003wk}.
The results of the calculation agree also with a simpler calculation
using the equivalent photon approximation in most cases. The exception
is the distribution of the transverse momentum of the pair.

In the lowest order calculation electrons and positrons have
an identical differential distribution. No asymmetry was also
seen in the experimental data, which has a large statistical error. 
Also the total cross section was found to be in agreement
with the lowest order calculation. No higher order reduction, as
predicted by the Bethe-Maximon theory was observed, which should be a
reduction of about 17\%. But this  is mainly due to the restriction on
the phase space by the experiment, where these corrections are not
expected.

\subsection{Bound-free pair production at RHIC
and the forthcoming LHC(Pb-Pb)}

The bound-free pair production (BFPP) 
(see the discussion in Sec.~\ref{ssec:boundfree})
is a process which is  
of practical importance for the operation of the LHC in the Pb-Pb mode. 
An electron-positron pair is produced, where the electron is produced
not as a free particle but into a bound atomic state of one of the
ions,
\BE
Z_1+Z_2 \rightarrow (Z_1+e^-)_{1s_{1/2},\cdots} + e^+ + Z_2.
\EE

As this changes the charge state of the nucleus, it is
lost from the beam in the collider.
Together with the electromagnetic dissociation of
the nuclei (see \cite{PRPLC.364.359,Baur:1998ay}  these two processes are the
dominant loss processes for heavy ion colliders. It has been
realized  \cite{Klein:2000ba}, (see also 
\cite{Brandt00,Jeanneret00}), that the BFPP process can also result 
in a localized beam-pipe heating: the atomic states are produced with 
a small perpendicular momentum of the order of $m$, and therefore
leads to a narrow singly-charged ion beam (The process of electromagnetic 
dissociation is  less severe in this respect as  the momentum transfer
to the ions is more spread out and the ions are not so strongly focussed
on a single spot). These beams with altered
magnetic rigidity will deposit their energy in a localized region of 
the beam pipe and  cause a localized heating. This can lead to
the quenching of the superconducting magnets. The energy deposited 
per unit time is of the order of Watts. This limits the luminosity of
the Pb-Pb collider at LHC \cite{Brandt00}.
The effect of BFPP is studied more quantitatively through simulations
of the ion showers in a dipole magnet in
\cite{Bruce:2006aa,Jowett:2005ci}.
These calculations are based on the cross section given in
\cite{Meier:2000ga}. In this reference, a comparison is also made with
the results of other calculations.
The authors of \cite{Bruce:2006aa} conclude that
``\ldots the secondary beam of ions 
emerging from each collision point 
at the nominal peak luminosity 
$L=1.0\times 10^{27}cm^{-2}s^{-1}$
is not likely to quench a dipole magnet''.

Due to the lower beam energy at RHIC,
the energy deposit there is smaller and therefore
not important as a limiting factor to the beam luminosity.
However, it can serve as a testing ground for the theoretical
numbers given in \cite{Meier:2000ga}.
This is especially important, as the experimental result with the
highest energies is from a fixed target measurement with an energy of
160 GeV per nucleon \cite{Krause98}, corresponding to a $\gamma=9.3$,
still far away from the situation at the LHC. 

First results of a measurement of ion beam losses due
to BFPP in Cu-Cu collisions at RHIC were reported in \cite{Jowett:2006cg}.
Due to the small number of events observed the authors were not able
to give a value for the cross section, but they conclude that within 
the uncertainties associated with the experimental conditions the data 
are consistent with the signal expected from BFPP. It is a valuable 
test of the ability to quantitatively predict the BFPP effect for the
LHC.

\subsection{Outlook on LHC}

In \cite{Hencken:2004td} also an outlook to the situation for the LHC
is given. One of the drawbacks of all 
possible measurements at high energies is
the fact that pair production is predominantly produced in the forward
direction, whereas the detectors are designed towards central
events. But most of the interesting effects, like Coulomb correction
and multiple pair production will occur at small transverse
momenta. In \cite{Hencken:2002an} it is discussed how the very inner
part of the detectors can be used for pair production detection. The
limits of the transverse momentum can be reduced to $p_t>2.6$MeV,
which is comparable to the rest mass. This is explored in more
detail in \cite{Hencken:2004td}. It is found that about 10\% of all
pairs produced together with nuclear excitation 
will come from multiple pair production. 

Another approach has been studied in \cite{Bocian:2004ev}.
Here the very forward CASTOR detector is used to measure pair
production. Calculations in lowest order are done and it is proposed
to use this as a luminosity measurement for the LHC. Of course the
accuracy of such a luminosity measurement relies on the precise 
theoretical knowledge of this
process especially in this region and how large Coulomb correction
contribute to it. In combination with the nuclear
excitation of the ions it would also allow to study pair production at
small impact parameters.

\section{Conclusion}
\label{section:conclusion}

In April 1990 a workshop took place in Brookhaven 
with the title  'Can RHIC be used to test QED?' \cite{Fatyga:1990ek}.
We think that after about 17 years the answer to this question is
'no'. However, many theorists were motivated to deal with this topic. 
The gradual progress which was sometimes quite tortuous, is 
described in this report.     

In this review we studied electron-positron pair production in heavy ion
collisions at relativistic energies. 
There were quite a few papers in the last years
approaching this topic from various aspects.  
Whereas the lowest order Born result was known for a long time,
higher order (Coulomb) corrections have been under debate
for the last decades. These corrections were studied
using various methods.
One of these methods  follow the approach of Bethe and Maximon,
which was used to study Coulomb corrections to the 
pair production in photon-nucleus interactions (the Bethe-Heitler
process). QED perturbation theory was also used and it was shown 
\cite{Phys.Rev.D57.4025} that they yield the same result as 
the approach of Bethe and Maximon.
Difficulties with obtaining analytical solutions appear when 
more than one photon are attached to each nucleus, and no firm results have
been obtained up to now. However, for the total cross section these
effects are small. 

It has been found in theoretical calculations that higher order effects
can be quite sizeable for collisions where small impact parameters are
selected. But an
experimental confirmation has not been made up to now.

The fields occurring in ultrarelativistic heavy ion collisions are 
very strong. So one may expect that some new kind of nonperturbative 
effects could arise, similar to the Schwinger mechanism
for static electric fields. But the fields 
in the relativistic heavy ion collision act only for a short time and
higher order perturbation theory is appropriate.
We also discuss the problem of the transition 
from slow to fast collisions. However, this chapter is far from 
being solved. 
A new higher order effect in relativistic heavy ion reactions
is the emission of multiple pairs in a single collision.
This is quite a spectacular effect and it would be of some interest 
to observe it experimentally in forthcoming experiments at the LHC.

In addition to the intrinsic interest of the theoretical study
of pair production at high energies there is the practical importance 
of these processes in the more general study of ultraperipheral 
processes. Pair production is a background process, e.g., 
to vector meson photoproduction.
The very large cross section for bound-free pair production
limits the maximum Pb-Pb luminosity at the forthcoming LHC.
Thus it is mandatory to understand these QED effects to a 
high degree of accuracy.

Finally, we review the experimental results on pair production at 
RHIC and their theoretical interpretation. Results
on the bound-free pair production mechanism obtained recently at RHIC
are also discussed. Agreement with theory is good, and we expect
a similar good agreement at the forthcoming LHC.
This good agreement is of course also the reason why electron and muon pairs 
produced in ultraperipheral collisions may well
be used  to monitor the luminosity at colliders.
This should not be surprising since QED is the best-tested theory we have.

\section*{Acknowledgements}

We are grateful to A.~Alscher, A.~Aste, C.~A.~Bertulani, U.~Dreyer,
S.~R.~Klein, H.~Meier, E.~A.~Kuraev, P.~Stagnioli, V.~Serbo for their
collaboration on various topics of this review. 
Furthermore we wish to acknowledge very interesting and helpful
dicussions with  many people, we would like to mention especially  
A.~Baltz, A.~Milstein, N.~N.~Nikolaev, W.~Scheid, G.~Soff, and 
T.~St\"ohlker.

%%%%%%%%%%%%%%%%%%%%%%%%%%%%%%%%%%%%%%%%%%%%%%%%%%%%%%%%%%%%%%%%%%%%%%
%\bibliography{pr}

\begin{thebibliography}{10}

\bibitem{Phys.Rev.76.790}
J.~S. Schwinger, Phys. Rev. {\bf 76},  790  (1949).

\bibitem{Phys.Rev.76.749}
R.~P. Feynman, Phys. Rev. {\bf 76},  749  (1949).

\bibitem{Phys.Rev.76.769}
R.~P. Feynman, Phys. Rev. {\bf 76},  769  (1949).

\bibitem{Schwinger:1958}
J.~S. Schwinger (Editor), {\em Selected Papers on QED} (Dover, New York, 1958).

\bibitem{Anderson:1933xx}
C.~D. Anderson, Phys. Rev. {\bf 43},  491  (1933).

\bibitem{Landau:1935aa}
L.~D. Landau and E.~M. Lifshitz, Phys. Z. Sowjet. {\bf 6},  244  (1934).

\bibitem{NUCIA.14.93}
G. Racah, Nuovo Cim. {\bf 14},  93  (1937).

\bibitem{PRPLC.15.181}
V.~M. Budnev, I.~F. Ginzburg, G.~V. Meledin, and V.~G. Serbo, Phys. Rept. {\bf
  15},  181  (1974).

\bibitem{Rafelski:1978}
J. Rafelski, L.~P. Fulcher, and A. Klein, Phys. Reports {\bf 38C},  227
  (1978).

\bibitem{GreinerStrongField}
W. Greiner, B. M{\"u}ller, and J. Rafelski, {\em Quantum Electrodynamics of
  Strong Fields} (Springer Verlag, Berlin, 1985).

\bibitem{Phys.Rev.82.664}
J.~S. Schwinger, Phys. Rev. {\bf 82},  664  (1951).

\bibitem{ItzyksonZ80}
C. Itzykson and J.-B. Zuber, {\em Quantum Field Theory} (McGraw-Hill,
  Singapore, 1980).

\bibitem{Ringwald:2003er}
A. Ringwald,  in {\em Electromagnetic probes of fundamental physics}, edited by
  W. Marciano and S. White (World Scientific Publishing Company, Singapore,
  2003), p.\ 67.

\bibitem{Ringwald:2001ib}
A. Ringwald, Phys. Lett. {\bf B510},  107  (2001).

\bibitem{Phys.Rev.D2.1191}
E. Brezin and C. Itzykson, Phys. Rev. {\bf D2},  1191  (1970).

\bibitem{Krauss:1997vr}
F. Krauss, M. Greiner, and G. Soff, Prog. Part. Nucl. Phys. {\bf 39},  503
  (1997).

\bibitem{Baur:1998ay}
G. Baur, K. Hencken, and D. Trautmann, J. Phys. {\bf G24},  1657  (1998).

\bibitem{PRPLC.364.359}
G. Baur {\it et~al.}, Phys. Rept. {\bf 364},  359  (2002).

\bibitem{Bertulani:2005ru}
C.~A. Bertulani, S.~R. Klein, and J. Nystrand, Ann. Rev. Nucl. Part. Sci. {\bf
  55},  271  (2005).

\bibitem{PRPLC.163.299}
C.~A. Bertulani and G. Baur, Phys. Rept. {\bf 163},  299  (1988).

\bibitem{Baltz:2007hw}
A. Baltz {\it et~al.}, Photoproduction at collider energies: From RHIC and HERA
  to the LHC, hep-ph/0702212, 2007.

\bibitem{ect07website}
Workshop on Photoproduction at collider energies: from RHIC and HERA to LHC,
  http://dde.web.cern.ch/dde/photoprod\_ect07/.

\bibitem{NUPHA.A729.787}
G. Baur {\it et~al.}, Nucl. Phys. {\bf A729},  787  (2003).

\bibitem{Bethe:1954bmd}
H.~A. Bethe and L.~C. Maximon, Phys. Rev. {\bf 93},  768  (1954).

\bibitem{PhysRev.93.788}
H. Davies, H.~A. Bethe, and L.~C. Maximon, Phys. Rev. {\bf 93},  788  (1954).

\bibitem{Landau:1986aa}
L.~D. Landau and E.~M. Lifschitz, {\em Quantenelektrodynamik}, No.~IV in {\em
  Lehrbuch der theoretischen Physik} (Akademie Verlag, Berlin, 1986).

\bibitem{Phys.Rev.D57.4025}
D. Ivanov and K. Melnikov, Phys. Rev. {\bf D57},  4025  (1998).

\bibitem{Meier:1997mp}
H. Meier, K. Hencken, D. Trautmann, and G. Baur, Eur. Phys. J. {\bf C2},  741
  (1998).

\bibitem{Phys.Rev.C58.1679}
A.~J. Baltz and L.~D. McLerran, Phys. Rev. {\bf C58},  1679  (1998).

\bibitem{nucl-th/0101024}
A.~J. Baltz, F. Gelis, L.~D. McLerran, and A. Peshier, Nucl. Phys. {\bf A695},
  395  (2001).

\bibitem{Phys.Rev.A57.1849}
B. Segev and J.~C. Wells, Phys. Rev. {\bf A57},  1849  (1998).

\bibitem{Phys.Rev.C59.2753}
B. Segev and J.~C. Wells, Phys. Rev. {\bf C59},  2753  (1999).

\bibitem{Aste:2001te}
A. Aste {\it et~al.}, Eur. Phys. J. {\bf C23},  545  (2002).

\bibitem{Phys.Rev.C71.024901}
A.~J. Baltz, Phys. Rev. {\bf C71},  024901  (2005).

\bibitem{Phys.Rev.A59.1223}
U. Eichmann, J. Reinhardt, S. Schramm, and W. Greiner, Phys. Rev. {\bf A59},
  1223  (1999).

\bibitem{Ivanov:1998ru}
D.~Y. Ivanov, A. Schiller, and V.~G. Serbo, Phys. Lett. {\bf B454},  155
  (1999).

\bibitem{Phys.Rev.A65.022101}
R.~N. Lee, A.~I. Milstein, and V.~G. Serbo, Phys. Rev. {\bf A65},  022101
  (2002).

\bibitem{Adams:2004rz}
J. Adams {\it et~al.}, Phys. Rev. {\bf C70},  031902 (R)  (2004).

\bibitem{Euler:2007}
http://www.euler-2007.ch.

\bibitem{Phys.Rev.A69.022708}
R.~N. Lee, A.~I. Milstein, and V.~M. Strakhovenko, Phys. Rev. {\bf A69},
  022708  (2004).

\bibitem{PHRVA.A50.3980}
A.~W. Aste, K. Hencken, D. Trautmann, and G. Baur, Phys. Rev. {\bf A50},  3980
  (1994).

\bibitem{Meier:2000ga}
H. Meier {\it et~al.}, Phys. Rev. {\bf A63},  032713  (2001).

\bibitem{Phys.Rev.173.1011}
S.~J. Brodsky and J. Gillespie, Phys. Rev. {\bf 173},  1011  (1968).

\bibitem{Bartos:2001jz}
E. Bartos, S.~R. Gevorkyan, E.~A. Kuraev, and N.~N. Nikolaev, Phys. Rev. {\bf
  A66},  042720  (2002).

\bibitem{Bartos:2004ss}
E. Bartos, S.~R. Gevorkyan, E.~A. Kuraev, and N.~N. Nikolaev, J. Exp. Theor.
  Phys. {\bf 100},  645  (2005).

\bibitem{Gevorkyan:2003dp}
S.~R. Gevorkyan and E.~A. Kuraev, J. Phys. {\bf G29},  1227  (2003).

\bibitem{Bartos:2002ii}
E. Bartos, S.~R. Gevorkyan, E.~A. Kuraev, and N.~N. Nikolaev, Phys. Lett. {\bf
  B538},  45  (2002).

\bibitem{Gevorkyan:2005ys}
S.~R. Gevorkyan and A.~V. Tarasov, Challenge of lepton pair production in
  peripheral collisions of relativistic ions, hep-ph/0512098, 2005.

\bibitem{Baltz:1991xx}
A.~J. Baltz, M.~J. Rhoades-Brown, and J. Weneser, Phys. Rev. A {\bf 44},  5569
  (1991).

\bibitem{AlderW66}
{\em Coulomb Excitation}, {\em Perspectives in Physics}, edited by K. Alder and
  A. Winther (Academic Press, New York, London, 1966).

\bibitem{Hencken:1994hp}
K. Hencken, D. Trautmann, and G. Baur, Phys. Rev. {\bf A51},  998  (1995).

\bibitem{BialynickiB75}
I. Bial{\-l}ynicki-Birula and Z. Bia{\-l}ynicki-Birula, {\em Quantum
  Electrodynamics} (Pergamon
%, ADDRESS, 
1975).

\bibitem{Lee:1999ey}
R.~N. Lee and A.~I. Milstein, Phys. Rev. {\bf A61},  032103  (2000).

\bibitem{Lee:2001aa}
R.~N. Lee and A.~I. Milstein, Phys. Rev. {\bf A64},  032106  (2001).

\bibitem{segev:1998ur}
B. Segev and J.~C. Wells, Phys. Rev. {\bf C59},  2753  (1999).

\bibitem{Abramowitz:1964aa}
M. Abramowitz and I.~A. Stegun, {\em Handbook of mathematical Functions}
  (Dover, New York, 1965).

\bibitem{Baltz:2004dz}
A.~J. Baltz, Phys. Rev. {\bf C71},  024901  (2005).

\bibitem{Baltz:2003dy}
A.~J. Baltz, Phys. Rev. {\bf C68},  034906  (2003).

\bibitem{Baltz:2005ay}
A.~J. Baltz, Acta Phys. Hung. {\bf A27},  323  (2006).

\bibitem{Baltz:2006mz}
A.~J. Baltz, Phys. Rev. {\bf C74},  054903  (2006).

\bibitem{Lee:2006ze}
R.~N. Lee and A.~I. Milstein, J. Exp. Theor. Phys. {\bf 104},  423  (2007).

\bibitem{Lee:2001ea}
R.~N. Lee, A.~I. Milstein, and V.~G. Serbo, Structure of the Coulomb and
  unitarity corrections to the cross-section of e+ e- pair production in
  ultrarelativistic nuclear collisions, 2001.

\bibitem{Hencken:1994my}
K. Hencken, D. Trautmann, and G. Baur, Phys. Rev. {\bf A51},  1874  (1995).

\bibitem{Guclu95}
M.~C. G{\"u\c cl\"u} {\it et~al.}, Phys. Rev. A {\bf 51},  1836  (1995).

\bibitem{Baur:1990za}
G. Baur, Phys. Rev. {\bf A42},  5736  (1990).

\bibitem{Glauber:1963tx}
R.~J. Glauber, Phys. Rev. {\bf 131},  2766  (1963).

\bibitem{RhoadesBrownW91}
M.~J. Rhoades-Brown and J. Weneser, Phys. Rev. A {\bf 44},  330  (1991).

\bibitem{BestGS92}
C. Best, W. Greiner, and G. Soff, Phys. Rev. A {\bf 46},  261  (1992).

\bibitem{Erlykin75a}
A.~D. Erlykin,  in {\em 14th Int. Cosmic Ray Conf., M{\"u}nchen, Conf. Papers},
%  edited by  (PUBLISHER, ADDRESS, 1975), 
Vol.~6, p.\ 1987.

\bibitem{Erlykin75b}
A.~D. Erlykin,  in {\em 14th Int. Cosmic Ray Conf., M{\"u}nchen, Conf. Papers},
%  edited by  (PUBLISHER, ADDRESS, 1975), 
Vol.~7, p.\ 2173.

\bibitem{Baltz:2001dp}
A.~J. Baltz, F. Gelis, L.~D. McLerran, and A. Peshier, Nucl. Phys. {\bf A695},
  395  (2001).

\bibitem{Alscher:1996gn}
A. Alscher, K. Hencken, D. Trautmann, and G. Baur, Phys. Rev. {\bf A55},  396
  (1997).

\bibitem{Hencken:1998hf}
K. Hencken, D. Trautmann, and G. Baur, Phys. Rev. {\bf C59},  841  (1999).

\bibitem{VaneDDD97}
C.~R. Vane {\it et~al.}, Phys. Rev. A {\bf 56},  3682  (1997).

\bibitem{Baltz:1996hr}
A.~J. Baltz, Phys. Rev. Lett. {\bf 78},  1231  (1997).

\bibitem{Eichler:2005xn}
J. Eichler, Phys. Lett. {\bf A347},  67  (2005).

\bibitem{Messiah:1964aa}
A. Messiah, {\em Quantum Mechanics, Vol. II} (Nort-Holland Publishing Company,
%  ADDRESS, 
1964).

\bibitem{BetheS77}
H.~A. Bethe and E.~E. Salpeter, {\em Quantum Mechanics of One- and Two-Electron
  Atoms} (Plenum, New York, 1977).

\bibitem{brabec:2000aa}
T. Brabec and F. Krausz, Rev. Mod. Phys. {\bf 72},  545  (2000).

\bibitem{PhysRevD.2.1191}
E. Brezin and C. Itzykson, Phys. Rev. D {\bf 2},  1191  (1970).

\bibitem{Merkel:2006aa}
A. Merkel, Science {\bf 313},  147  (2006).

\bibitem{mourou:309}
G.~A. Mourou, T. Tajima, and S.~V. Bulanov, Reviews of Modern Physics {\bf 78},
   309  (2006).

\bibitem{Ringwald:2004aa}
Jena, Mai 2004 Kolloqium, http://desy.de/~ringwald/xfel/talks.

\bibitem{Alkofer:2001ik}
R. Alkofer {\it et~al.}, Phys. Rev. Lett. {\bf 87},  193902  (2001).

\bibitem{Eichler90}
J. Eichler, Physics Report {\bf 193},  165  (1990).

\bibitem{Hencken:2004td}
K. Hencken, G. Baur, and D. Trautmann, Phys. Rev. {\bf C69},  054902  (2004).

\bibitem{Hencken:1993cf}
K. Hencken, D. Trautmann, and G. Baur, Phys. Rev. {\bf A49},  1584  (1994).

\bibitem{Morozov:2003wk}
V.~B. Morozov, Electron positron production in ultra-peripheral heavy-ion
  collisions with the STAR experiment, nucl-ex/0403002, 2003.

\bibitem{Klein:2000ba}
S.~R. Klein, Nucl. Instrum. Meth. {\bf A459},  51  (2001).

\bibitem{Brandt00}
D. Brandt, Review of the LHC Ion Report, LHC Project Report 450, 2000.

\bibitem{Jeanneret00}
J.~B. Jeanneret, Electron capture in Pb-Pb collisions and quench limit, Beam
  Physics Note 41, 2000.

\bibitem{Bruce:2006aa}
R. Bruce, S. Gilardone, and J.~M. Jowett, bound free pair production loss and
  quench limit for LHC magnets, LHC Note 379, 2006.

\bibitem{Jowett:2005ci}
J.~M. Jowett, R. Bruce, and S.~S. Gilardoni, Luminosity limit from bound-free
  pair production in the LHC, Prepared for Particle Accelerator Conference (PAC
  05), Knoxville, Tennessee, 16-20 May 2005.

\bibitem{Krause98}
H.~F. Krause {\it et~al.}, Phys. Rev. Lett. {\bf 80},  1190  (1998).

\bibitem{Jowett:2006cg}
J.~M. Jowett {\it et~al.}, Measurement of ion beam losses due to bound-free
  pair production in RHIC, Prepared for European Particle Accelerator
  Conference (EPAC 06), Edinburgh, Scotland, 26-30 Jun 2006, see
  http://accelconf.web.cern.ch/AccelConf/e06/PAPERS/MOPLS010.PDF.

\bibitem{Hencken:2002an}
K. Hencken, Y. Kharlov, and S. Sadovsky, Ultraperipheral Trigger in ALICE,
  ALICE Internal Note ALICE-INT-2002-11, 2002.

\bibitem{Bocian:2004ev}
D. Bocian and K. Piotrzkowski, Acta Phys. Polon. {\bf B35},  2417  (2004).

\bibitem{Fatyga:1990ek}
M. Fatyga, M. Rhoades-Brown, and M. Tannenbaum, Can RHIC be used to test QED:
  Workshop summary, Workshop ``Can RHIC be used to test QED?'', Upton, N.Y.,
  Apr 20-21, 1990, BNL 52247 Formal Report.

\end{thebibliography}
%\bibliographystyle{prsty}
\end{document}